\PassOptionsToPackage{pdfpagelabels=false}{hyperref}
\documentclass[fleqn,usenatbib,usedcolumn]{mnras}
\usepackage[british]{babel}             
\usepackage{txfonts}                  

\usepackage{graphicx}	
\usepackage{hyperref}	
\usepackage{soul}
\usepackage{bm}

\newcommand*\rad{~rad\,m$^{-2}$}

\newcommand*\radsq{~rad$^2$\,m$^{-4}$}

\newcommand*\rms{$\Delta{\rm RM_{rms}}$}
\newcommand*\dRM{$\Delta{\rm RM}$}
\newcommand*\dRMsq{$(\Delta{\rm RM})^2$}

\DeclareRobustCommand{\ion}[2]{%
\relax\ifmmode
\ifx\testbx\f@series
{\mathbf{#1\,\mathsc{#2}}}\else
{\mathrm{#1\,\mathsc{#2}}}\fi
\else\textup{#1\,{\mdseries\textsc{#2}}}%
\fi}

\title[IGMF limits with LOFAR]{New constraints on the magnetization of the cosmic web using LOFAR Faraday rotation observations}
\author[O'Sullivan et al.]{S.~P.~O'Sullivan$^{1,2}$\thanks{E-mail: shane.osullivan@dcu.ie}, M.~Br\"uggen$^{1}$, F.~Vazza$^{3,1,4}$, 
E.~Carretti$^{4}$, N.~T.~Locatelli$^{3,4}$, C.~Stuardi$^{3,4}$, \and V.~Vacca$^{5}$, T.~Vernstrom$^{6}$, G.~Heald$^{6}$, C.~Horellou$^{7}$, T.~W.~Shimwell$^{8,9}$, M.~J.~Hardcastle$^{10}$, \and C.~Tasse$^{11}$, H.~R\"ottgering$^{9}$\\ 
$^{1}$ Hamburger Sternwarte, Universit\"at Hamburg, Gojenbergsweg 112, Hamburg 21029, Germany.\\
$^{2}$ School of Physical Sciences and Centre for Astrophysics \& Relativity, Dublin City University, Glasnevin, D09 W6Y4, Ireland.\\
$^{3}$ Dipartimento di Fisica e Astronomia, Universit\'a di Bologna, Via Gobetti 92/3, 40121, Bologna, Italy.\\
$^{4}$ Istituto di Radio Astronomia, INAF, Via Gobetti 101, 40121 Bologna, Italy.\\
$^{5}$ INAF - Osservatorio Astronomico di Cagliari, Via della Scienza 5, I-09047 Selargius (CA), Italy. \\
$^{6}$ CSIRO Astronomy and Space Science, PO Box 1130, Bentley WA 6102, Australia.\\
$^{7}$ Dept. of Space, Earth and Environment, Chalmers University of Technology, Onsala Space Observatory, SE-43992 Onsala, Sweden. \\
$^{8}$ ASTRON, the Netherlands Institute for Radio Astronomy, Postbus 2, 7990 AA, Dwingeloo, The Netherlands. \\
$^{9}$ Leiden Observatory, Leiden University, PO Box 9513, NL-2300 RA Leiden, The Netherlands. \\
$^{10}$ Centre for Astrophysics Research, School of Physics, Astronomy and Mathematics, University of Hertfordshire, College Lane, Hatfield AL10 9AB, UK.\\
$^{11}$ GEPI \& USN, Observatoire de Paris, Universit\'e PSL, CNRS, 5 Place Jules Janssen, 92190 Meudon, France. \\
}

\begin{document}

\date{Accepted 2020 May 13. Received 2020 May 13; in original form 2020 February 07.}

\pagerange{\pageref{firstpage}--\pageref{lastpage}} \pubyear{2019}

\maketitle

\label{firstpage}

\begin{abstract} 
Measuring the properties of extragalactic magnetic fields through the effect of Faraday 
rotation provides a means to understand the origin and evolution of cosmic magnetism. 
Here we use data from the LOFAR Two-Metre Sky Survey (LoTSS)  
to calculate the Faraday rotation measure (RM) of close pairs of extragalactic radio sources. 
By considering the RM difference (\dRM) between physical pairs (e.g.~double-lobed radio galaxies) 
and non-physical pairs (i.e.~close projected sources on the sky), 
we statistically isolate the contribution of extragalactic magnetic fields 
to \dRM~along the line of sight between non-physical pairs. 
From our analysis, we find no significant difference between the \dRM~distributions of the 
physical and non-physical pairs, limiting the excess Faraday 
rotation contribution to $< 1.9$\rad~($\sim$$95\%$ confidence). 
We use this limit with a simple model of an inhomogeneous universe to place an upper limit 
of 4 nG on the cosmological co-moving magnetic field strength on Mpc scales. We also compare the 
RM data with a more realistic suite of cosmological MHD simulations, that explore different 
magnetogenesis scenarios. Both magnetization of the large scale structure by astrophysical processes 
such as galactic and AGN outflows, and simple primordial scenarios with seed magnetic field 
strengths $< 0.5$~nG cannot be rejected by the current data; while stronger primordial fields or 
models with dynamo amplification in filaments are disfavoured. 
\end{abstract}


\section{Introduction}\label{sec:intro}
Uncovering the origin and understanding the evolution of cosmic magnetic fields is one of the key science 
goals for present and future radio telescopes \citep[e.g.][]{gaenslerSKA2004,akahori2018}. 
In addition to understanding the influence of magnetic fields on a range of different astrophysical scales,
these studies can provide important information on the physics of the early Universe \citep{widrow2012,jedamzik2020}. 
In particular, detecting the presence of magnetic fields in cosmic filaments and voids can provide key constraints on 
magnetogenesis scenarios \citep{durrerneronov2013,kandu2015}, mainly because they are not as strongly modified as 
the magnetic fields in galaxies and galaxy cluster environments. 
Direct detection of the non-thermal synchrotron emission associated with fields in cosmic filaments 
may be possible \citep[][]{vacca2018,vazza2019}, while an alternative approach is to use the Faraday rotation 
of linearly polarized radio sources to measure the field strength in thermal magnetized 
plasma along the line of sight \citep[][]{kronbergperry1982,orenwolfe1995,kolatt1998,stasyszyn2010,akahori2014}. 
This approach should also be possible in future large surveys of Fast Radio Bursts (FRB), provided that thousands of 
FRB rotation measures will be available \citep[e.g.][]{hackstein2019}. 

The magnetic field properties of galaxies and the intergalactic medium in groups and clusters 
of galaxies are well studied \citep[e.g.][]{carillitaylor2002,laing2008,beck2015,vanweeren2019}. 
However, the magnetic field properties of the pristine gas far outside 
galaxy clusters in filaments and voids are poorly constrained, with upper limits ranging from tens of 
nano-gauss \citep{ravi2016,vernstrom2019} and less \citep{blasi1999,PLANCK2015,pshirkov2016,hackstein2016,bray2018} 
to lower limits of $\sim$10$^{-17}$~G \citep{neronovvovk2010,tavecchio2011,dermer2011,dolag2011,taylor2011}. 
Improving our understanding of the strength and morphology of these fields will help to discriminate 
between competing models for the origin of cosmic magnetism, such as a primordial origin \citep{grasso2001,widrow2002,kulsrudzweibel2008}
or at later times through AGN and/or galactic outflows \citep{zweibel1997,furlanettoloeb2001,widrow2002,beck2013}. 
Most notably, the fall off in field strength with distance from dense regions of the Universe is less steep in the case of a 
primordial origin compared to a later `magnetic pollution' through outflows \citep[][]{donnert2009,vazza2017}. 

Constraining the magnetization of the Universe on large scales can also help test models of dark matter. 
For example, axion-like particles (ALPs) are a promising dark matter candidate \citep[][]{1988PhRvD..37.1237R,2003JCAP...05..005C}, 
which are predicted to oscillate into high-energy photons (and back) in the presence of background magnetic fields 
\citep[][]{2012PhRvD..86g5024H}.  
Photon-ALP oscillations are estimated to be possible on scales of a few Mpc 
in the presence of magnetic field strengths ranging from $\sim$1 to $10 ~\rm nG$ \citep[][]{2017PhRvL.119j1101M}.

The focus of this paper is on using the Faraday rotation measure (RM) of a sample of extragalactic radio 
sources to constrain the properties of the intergalactic magnetic field (IGMF) on large scales. 
This approach probes the thermal gas density-weighted field strength along the line of sight, where  
\begin{equation}
{\rm RM}_{[{\rm rad~m}^{-2}]} = 0.812 \int_{\rm source}^{\rm telescope} n_{e\,\,[{\rm cm}^{-3}]} \,\, B_{||\,\,[{\rm \mu G}]} \,\, dl_{\,\,[\rm{pc}]} 
\label{rmeqn}
\end{equation}
with $B_{||}$ representing the line-of-sight magnetic field strength, $n_e$ the free electron number density, and $l$ the path length 
through the magnetoionic medium.
This is complementary to other radio studies which attempt to detect the faint synchrotron 
emission from relativistic electrons in the cosmic web between clusters of galaxies 
\citep[e.g.][]{brown2017,vernstrom2017,vacca2018,botteon2018,govoni2019}. 

In order to assess the contribution of the IGMF to the RM, we need to study the contributions 
to the RM along the entire line of sight. For a statistical sample of background polarized radio 
sources, we are primarily concerned with the RM variance generated 
from extragalactic Faraday rotating material ($\sigma_{\rm RM,ex}^2$) that can be local or 
internal to the radio source itself or from the intergalactic medium on large scales. Furthermore, 
there is a large contribution from the interstellar medium (ISM) of the Milky Way ($\sigma_{\rm RM,MW}^2$), 
and a small contribution from the time-variable RM of the Earth's ionosphere 
($\sigma_{\rm RM,ion}^2$), in addition to measurement errors ($\sigma_{\rm RM,err}^2$). 
The total RM variance is then 
\begin{equation}\label{eqn:rmvariance}
\sigma_{\rm RM}^2 = \sigma_{\rm RM,ex}^2 + \sigma_{\rm RM,MW}^2 + \sigma_{\rm RM,ion}^2 + \sigma_{\rm RM,err}^2 \, .
\end{equation}

The majority of recent investigations of RM variance have been conducted at 1.4~GHz, mainly due to 
the catalog of 37,543 RMs produced from the NRAO VLA Sky Survey data \citep[NVSS;][]{condon1998,taylor2009}. 
Most investigations have used this catalog to study the properties of the Milky Way \citep[e.g.][]{harveysmith2011,stil2011,oppermann2012,purcell2015,hutsch2019}. 
However, \cite{schnitzeler2010} and \cite{oppermann2015} modelled both the Galactic and extragalactic RM variance and found a best-fitting 
extragalactic RM variance of $\sim$7\rad. Recently, \cite{vernstrom2019} conducted an innovative study 
of close pairs of extragalactic RMs to isolate an extragalactic RM variance of $\sim$5 to 10\rad. 
The RM variance local to radio sources has been well studied for individual objects, typically in groups or 
clusters of galaxies where the hot, magnetized intracluster gas can dominate the RM variance \citep[e.g.][]{laing2008}. 
However, the contribution of Faraday rotating material directly related to the radio sources themselves 
can be significant in some cases \citep[e.g.][]{rudnickblundell2004,osullivan2013,anderson2018,banfield2019,knuettel2019}. 
Importantly, since the RM 
variance local to radio sources can vary from 10's to 1000's of\rad, isolating a population of low RM variance 
sources is a key objective for experiments that aim to probe foreground RM screens with high precision \citep{rudnick2019}. 
The ionospheric RM must also be considered \citep{sotomayor2013} since the typical contribution is of $O(1$\rad$)$, 
which is similar to or larger than the signal from the IGMF that we want to probe. 

In this paper, we present an RM study in quite a different regime for Faraday rotation, using the Low Frequency Array 
\citep[LOFAR;][]{vanhaarlem2013} at 144~MHz. In particular, we use data from the ongoing LOFAR Two-Metre 
Sky Survey \citep[LoTSS;][]{shimwell2019} from 120 to 168~MHz. This provides a wavelength-squared coverage 
more than 600 times that of the NVSS. Since the accuracy with which one can measure Faraday rotation depends on the 
wavelength-squared coverage, the 
advantage of RM studies at m-wavelengths is a dramatically higher 
precision on individual RM measurements \citep{osullivanlenc2018,vaneck2018,neld2018}. 
However, the effect of Faraday depolarization increases substantially at long wavelengths, 
and while this provides important information on the properties of the magnetic field, it also 
means that a smaller fraction of radio sources are polarized at a level which can be detected \citep[e.g.][]{farnsworth2011}. 
This means that to reach a comparable polarized source density on the sky, observations at metre 
wavelengths need to be much deeper than cm-wavelengths \citep[][]{osullivan2018b}. 
To date, the majority of polarized sources at m-wavelengths have been large FRII radio galaxies \citep[e.g.][]{vaneck2018}, 
whose polarized hotspots and/or outer lobe regions extend well beyond the host galaxy environment, making them 
excellent probes of the IGMF and ideal for this project. 

In studying the extragalactic RM with these data, we follow the strategy of \cite{vernstrom2019}, hereafter V19, 
of taking the RM difference between close pairs ($<20$~arcmin) and then separating the samples into physical and 
non-physical (or random) pairs. The physical pairs are typically double-lobed radio galaxies, while the non-physical 
pairs are sources that are close in projection on the sky but physically located at different redshifts. 
The key insight upon which this experiment is based is that the RM variance due to the Milky Way should vary with pair angular 
separation in a similar manner for the physical and non-physical pairs, while the extragalactic RM variance 
due to the IGMF is expected to be larger for the non-physical pairs because of the much larger cosmic 
separation along the line of sight. 

In Section~\ref{sec:data}, we describe the observational data and our classification scheme. 
The observational results are presented in Section~\ref{sec:results}. Two approaches to 
placing model limits on intergalactic magnetic fields are described in Section~\ref{sec:sims}, 
while a discussion of the implications are given Section~\ref{sec:discussion}, followed by the conclusions in Section~\ref{sec:conclusions}. 
Throughout this paper, we assume a $\Lambda$CDM cosmology with 
H$_0 = 67.8$ km s$^{-1}$ Mpc$^{-1}$, $\Omega_M=0.308$ and $\Omega_{\Lambda}=0.692$ \citep{planck2016xiii}.

\begin{figure*}
\includegraphics[width=17.5cm,clip=true,trim=0.0cm 0.0cm 0.0cm 0.0cm]{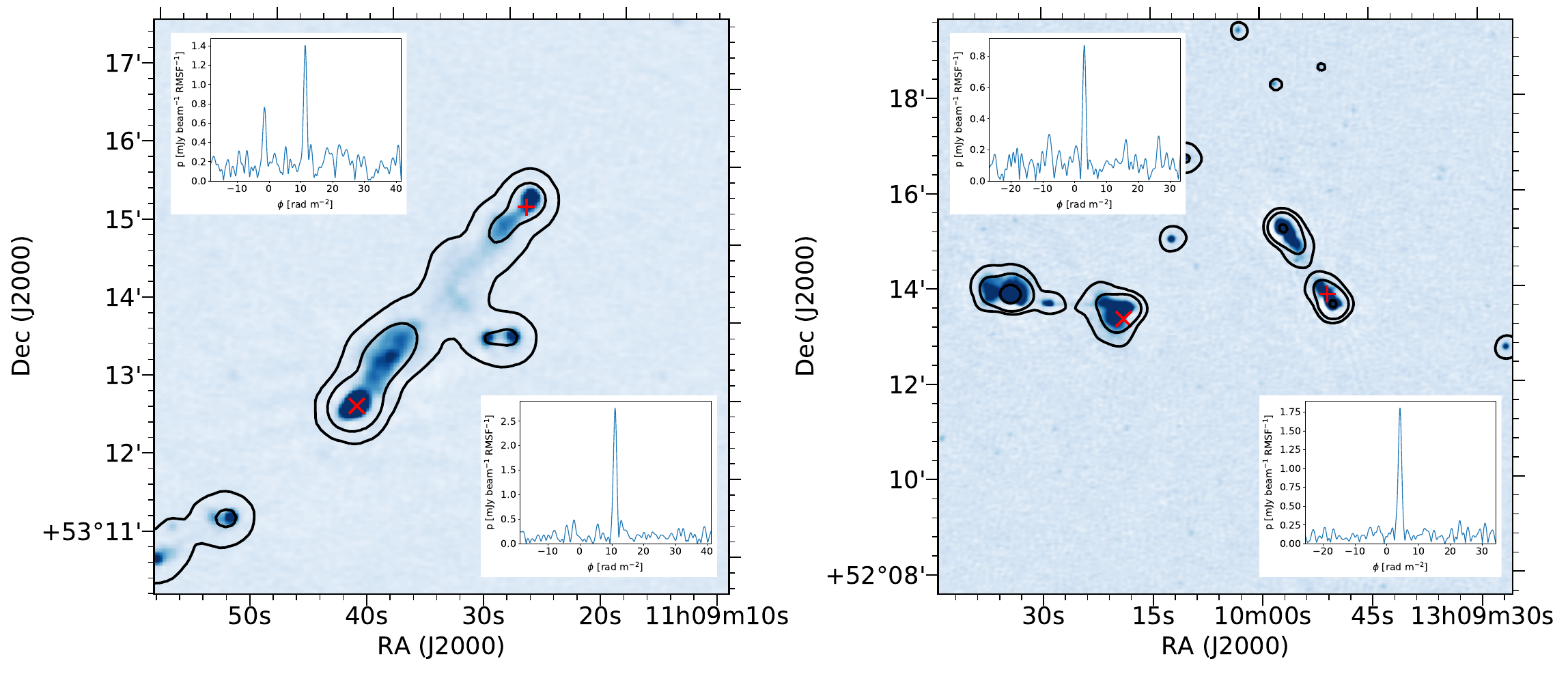}
\caption{Example of a physical RM pair (left) and a non-physical, random RM pair (right). 
The random RM pair on the right is composed of two double-lobed radio galaxies, for which linear 
polarization is only detected in one lobe of each. 
The cross and plus symbols represent the location of the peak polarized intensity from which the 
RM value is taken. The contours outline the total intensity emission at an angular resolution of 20~arcsec, which 
is the same resolution as the corresponding polarization data. 
The colorscale shows the 6~arcsec total intensity emission, used for the classification into physical 
and random pairs. 
The insets show the Faraday spectra from the location of the cross and plus symbols in the top left 
and bottom right, respectively. }\label{fig:pairs}
\end{figure*}

\section{Data}\label{sec:data}
The LoTSS is an ongoing survey of the northern sky with 
the LOFAR High Band Antennas, producing total intensity images and catalogs with an 
angular resolution of $\sim$6 arcsecond at 144~MHz \citep{shimwell2019}. 
From the second data release (DR2) survey pipeline (Tasse et al.~2020, in prep),  
polarization data products (Stokes $Q$, $U$ and $V$) are also being produced at an 
angular resolution of 20 arcseconds. In this work we make use of the Stokes $Q$ and $U$ data 
to find linearly polarized radio sources and derive their Faraday rotation measure (RM) values. 

The polarization data used here has a frequency range of 120 to 168~MHz with a channel bandwidth of 97.6~kHz. 
In order to efficiently find linearly polarized radio sources, we employ the technique of RM synthesis \citep{burn1966,bdb2005} 
where one takes a Fourier transform of the complex linear polarization vector, $\bm{P}(\lambda^2)$, defined as 
\begin{equation}
\bm{P}(\lambda^2)=\int_{-\infty}^{\infty} \bm{F}(\phi) \, e^{2i\phi\lambda^2} \, d\phi,
\end{equation}
to obtain the Faraday dispersion function, $\bm{F}(\phi)$, which provides the distribution of polarized emission 
as a function of Faraday depth ($\phi$) along the line of sight. In our case, the Faraday depth of the peak of $|\bm{F}(\phi)|$ 
is taken as the RM of the source. 
The LoTSS data provides an RM resolution of $\sim$1.15\rad~with a maximum scale of $\sim$1.0\rad~(i.e.~no 
sensitivity to resolved Faraday depth structures), and a maximum $|$RM$|$ of $\sim$170\rad~(up to 
$\sim$450\rad~with half the sensitivity). 
The time-variable absolute ionospheric RM correction was applied using {\sc rmextract}\footnote{https://github.com/lofar-astron/RMextract} 
as part of the standard {\sc prefactor} pipeline \citep[e.g.][]{degasperin2019}. 
Residual ionospheric RM correction errors of $\sim$0.1 to 0.3\rad~are expected across a single 8 hr observation \citep{sotomayor2013}.

The polarization catalog is produced by running RM synthesis\footnote{https://github.com/mrbell/pyrmsynth} 
on the Stokes $Q$ and $U$ images for each survey pointing out to a radius of 2 degrees from the pointing 
centre. 
The catalog used here is produced from 268 survey pointings which, considering the large overlap between 
adjacent pointings, covers a sky area of $\sim$2000 square degrees. The pointings used were not from a 
single contiguous sky area but were chosen from DR2 pointings that had been processed up to 2019 May 1. 
The pointings come from two (partially covered) areas of the sky, from RA of 7 to 19 hrs with Dec from 25 to 
70$^\circ$, and RA of 23 to 3 hrs with Dec from 10 to 40$^\circ$. 
The large overlap between pointings means that the same polarized sources are often found in multiple pointings. 
These duplicate sources were identified and only the source closest to a pointing centre was retained. 
The Faraday depth range was limited to $\pm$120\rad~with a sampling of 0.3\rad, mainly due 
to computer processing and storage limitations. Searching over a larger Faraday depth range (with a finer 
frequency channelisation) will be required to find sources in regions of the sky with high mean RM values, 
such as at low Galactic latitudes. A polarized source was cataloged if the peak in the 
Faraday dispersion function (FDF) was greater than 8 times the average noise in $Q$ and $U$ (i.e.~$\sigma_{QU}$, 
calculated from the rms of the wings of the real and imaginary parts of the FDF at $>$100\rad). 
For an $8\sigma_{QU}$ limit we expect a false detection rate of $\sim$$10^{-4}$, 
compared to $5\sigma_{QU}$ where it may be as high as $\sim$4\% \citep{george2012}. 
The $Q$ and $U$ frequency spectra were extracted at the source location and RM synthesis\footnote{https://github.com/CIRADA-Tools/RM} 
was applied with a finer sampling of 0.1\rad. The catalogued RM value was obtained from fitting a parabola to 
the amplitude of the complex FDF. 
The error in each RM value was calculated in the standard way \citep[e.g.][]{bdb2005} as the 
RM resolution divided by twice the signal to noise (i.e.~in our case this is $\sim0.58\sigma_{QU}/P$), 
where $P$ is the peak polarized intensity in the FDF after correction for the polarization bias 
following \cite{george2012}. 
Polarized sources in the Faraday depth range of $-3$ to $+1$\rad~were mainly discarded due to the 
presence of substantial contamination from instrumental polarization, which is shifted from 0\rad~by 
the ionospheric RM correction. 
The focus on this work is to obtain an initial catalog of close RM pairs. 
A more complete LOFAR RM catalog is under construction with more uniform sky coverage, in addition 
to the inclusion of sources without a close RM pair (O'Sullivan et al.~2020, in preparation). 

\subsection{Classification of RM pairs}
To obtain an initial sample of LOFAR RM pairs, we cross-matched the preliminary LOFAR RM catalog 
($\sim$1000 candidate polarized sources over $\sim$2000 sq.~deg.) with itself, 
including only pairs with separations $\leq20$ arcminutes. After removing self-matches, and limiting 
the minimum separation to 0.33 arcmin (i.e.~the angular resolution of the data of 20~arcsec), 
in addition to further quality assurance checks, we were left with 349 pairs. 
This matches the approach of V19, except for the minimum separation, which was limited to 1.5~arcmin in their study. 
All LOFAR pairs were restricted to come from the same pointing to avoid the systematic RM errors introduced 
by the different ionosphere corrections for different pointings. 
In fact, taking the RM difference between 
sources within the same pointing (as we describe later) removes the majority of the residual RM variance from the 
ionospheric RM correction, modulo direction-dependent variations on scales $<20$~arcmin \citep{degasperinionosphere}, 
leaving mainly the measurement errors from the observational noise remaining. 
This means there is a more precise measurement of the RM difference between pairs compared 
to the individual uncertainty on any one RM measurement. 

Visual inspection was used to separate sources into physical pairs (part of the same extragalactic 
radio source, e.g.~two lobes) and non-physical, random pairs (i.e.~physically unrelated sources projected close to 
each other on the sky). Classification of sources into physical and random pairs was done by S.~P.~O'Sullivan. 
This classification task was straightforward due to the high fidelity 
LoTSS Stokes $I$ images available at both 20 and 6 arcsecond resolution. 
All pairs are at Galactic latitude $|b|>20$~degrees, with no obvious clustering of physical or random 
pairs in particular parts of the sky. A Kolmogorov-Smirnov (KS) test provides no evidence for the two samples having a 
different underlying distribution in Galactic latitude ($p=0.2$).
The highest number of pairs for a single catalog source is 4, with a median of 1. 
Examples of physical and random pairs are shown in Fig.~\ref{fig:pairs}. 
Approximately 40\% of the random pairs have a compact source in the pair, while 
the resolved physical pairs are exclusively double-lobed radio galaxies. 

\subsection{The RM difference in pairs}
We are interested in investigating the difference in RM ($\Delta{\rm RM}={\rm RM}_1 - {\rm RM}_2$) 
between pairs of sources (i.e.~physical or random), in addition to the behaviour as a function of the 
angular separation ($\Delta\theta$). 
Since the RM difference can be positive or negative, we expect a mean \dRM~of zero for 
large samples. Therefore, 
we calculate the root-mean-squared (rms) in \dRM~as 
\begin{equation}
\Delta{\rm RM_{rms}} \equiv \sqrt{\langle (\Delta {\rm RM})^2 \rangle} = \sqrt{ \frac{1}{N}\sum_i{({\rm RM_1} - {\rm RM_2})^2_i}} \, .
\end{equation}

The RM variance contributed by the measurement errors ($\sigma_{\rm RM,err}^2$ in Eqn.~\ref{eqn:rmvariance}) 
can be subtracted from the total variance to analyse the astrophysical signal. 
We calculate this term from the quadrature sum of the errors on the individual RM measurements. 
Unless otherwise stated, the variance from measurement errors have been subtracted from the quoted \rms~values. 
For small samples or in the presence of outliers, the median can be a more robust statistic. 
Therefore, in our analysis we also consider the median of 
the absolute values of the RM difference (i.e.~ $|\Delta{\rm RM}|_{\rm median}$). 

\begin{figure}
\includegraphics[width=8.2cm,clip=true,trim=0.4cm 0.6cm 0.0cm 0.0cm]{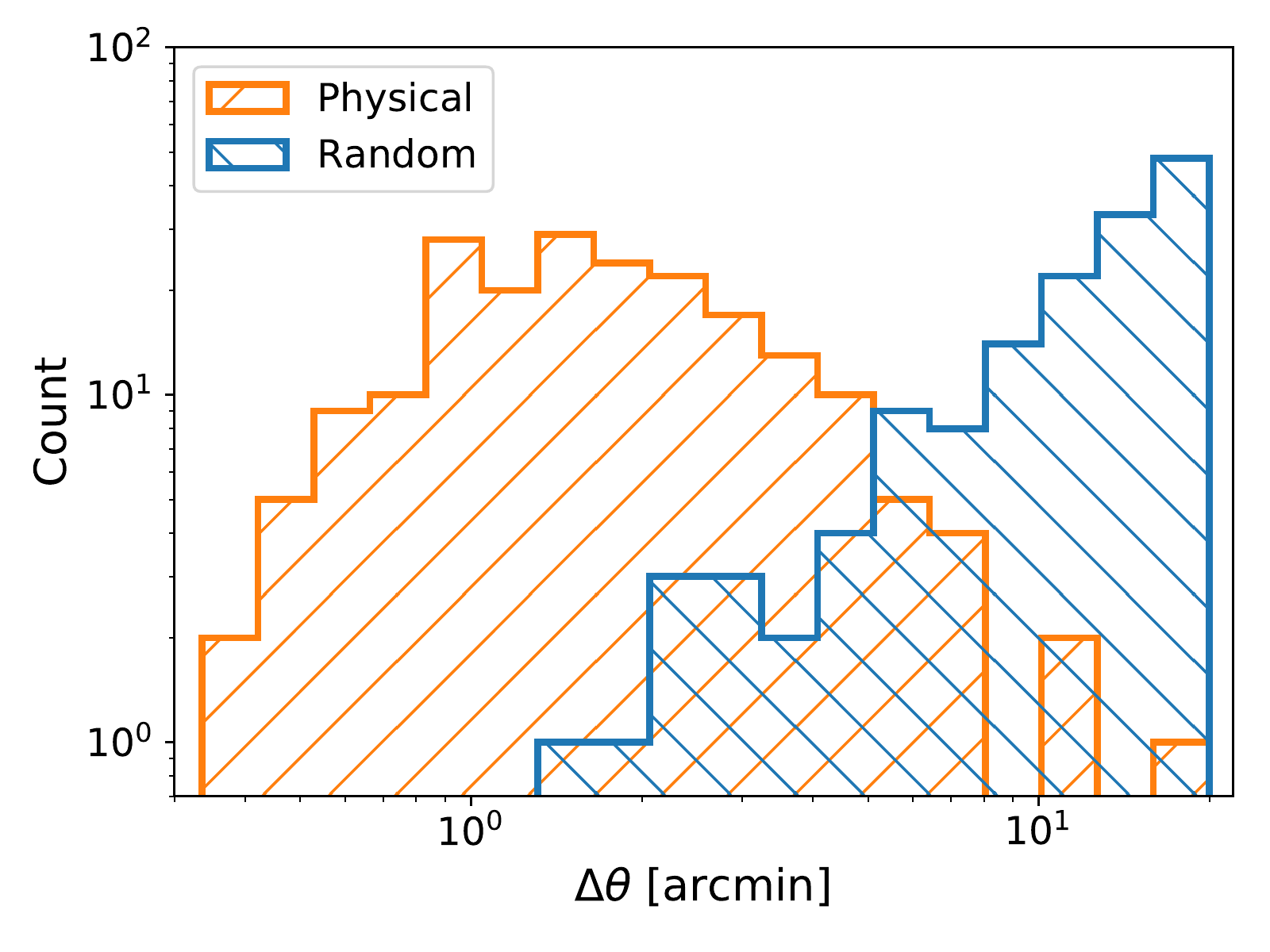}
\caption{Histogram of the angular separation ($\Delta\theta$, in units of arcminutes) of random (blue) and physical (orange) pairs. 
While the random pairs are typically found at larger angular separations, there is a significant overlap 
for random and physical pairs in the range of $\Delta\theta$ from 2 to 10~arcmin. }\label{fig:ang_sep}
\end{figure}

\section{Results}\label{sec:results}

In Fig.~\ref{fig:ang_sep}, we show the histogram of angular separations ($\Delta\theta$) for our sample of 148
random pairs (RPs) and 201 physical pairs (PPs). In both cases, we limited the maximum angular 
separation to 20~arcmin, with the RPs extending down to $\sim$1.5~arcmin and the PPs limited to the lower cut-off 
of 0.33~arcmin (i.e.~the angular resolution). The limit of 20~arcmin was chosen because there are very few PPs above this separation. 
There is a clear difference in that the PPs are more often found at smaller angular separations (mean of 2~arcmin) 
than the random pairs (mean of 12~arcmin). This is expected since the PPs are limited to the 
linear size of the radio source, while the RPs have no such restriction. 
Since we want to compare the RPs and PPs, we are mainly interested in the
region of significant overlap in angular separation between the two samples (in order to account 
for the Milky Way contribution). The overlap region we define here 
is from $\sim$2 to $\sim$10~arcmin, with 42 RPs and 75 PPs in this region (Table~\ref{tab:basic}). 
For context, V19 found 317 PPs and 5111 RPs on angular scales from 1.5 to 20'. 
They chose an overlap region of 3 to 11~arcmin, which contained 158 RPs and 208 PPs. 
Although V19 had significantly more sources, our measurement errors are much lower, 
such that both experiments have comparable statistical power. 
The individual RM values for each pair are provided in Table~\ref{tab:data}. 

\begin{figure}
\includegraphics[width=8.2cm,clip=true,trim=0.4cm 0.5cm 0.0cm 0.0cm]{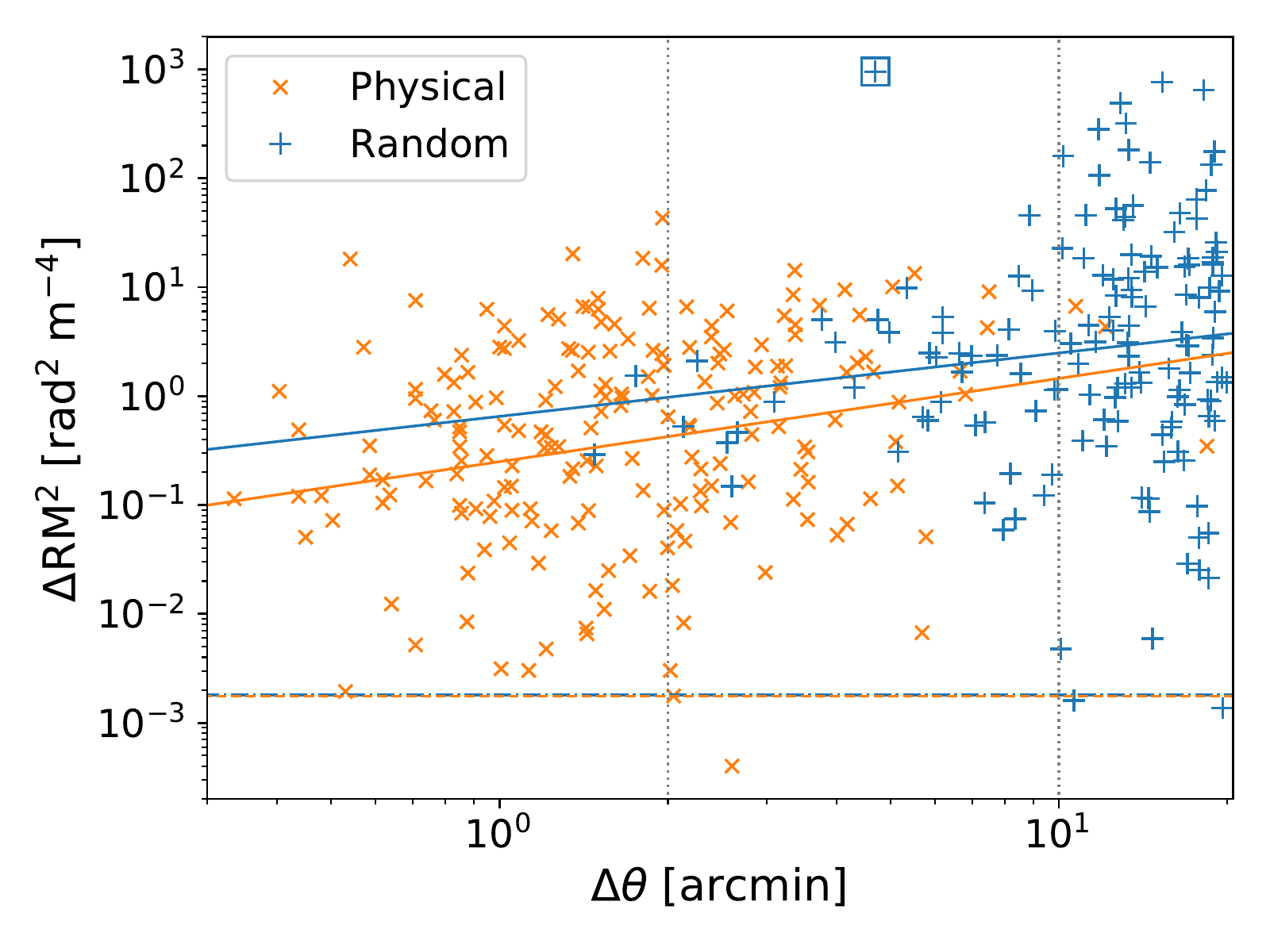}
\caption{Scatter plot of the squared difference in RM, \dRMsq~in units of \radsq, between pairs of radio sources versus the 
angular separation ($\Delta\theta$, in units of arcminutes). Physical pairs are shown as orange cross symbols while the 
random pairs are shown as blue plus symbols (with the outlier highlighted by a box). 
Power-law fits to the physical and random pair 
data are shown as solid orange and blue lines, respectively. The small, constant contributions to 
\dRMsq~from the measurement errors are shown for physical and random pairs as 
horizontal orange dashed and blue dot-dashed lines, respectively. 
The gray dotted vertical lines bound the overlap region of 2 to 10~arcmin. 
}\label{fig:dRMsq_sep}
\end{figure}

Fig.~\ref{fig:dRMsq_sep} shows the individual values of \dRMsq~and $\Delta\theta$ for 
each source pair, with the RPs indicated by plus symbols and the PPs indicated by cross symbols. 
The mean RM error for our sample is 0.026\rad, and has a small contribution to the overall variance. 
The variance added by the measurement errors (i.e.~$\sigma_{\rm RM,err}^2$ in Eqn.~\ref{eqn:rmvariance}) 
for physical and random pairs as a function of angular size is 
approximately constant, and shown in Fig.~\ref{fig:dRMsq_sep} as dashed and dot-dashed lines with 
values of $\sim$0.0018\radsq. 

\begin{figure}
\includegraphics[width=8.2cm,clip=true,trim=0.4cm 0.5cm 0.0cm 0.0cm]{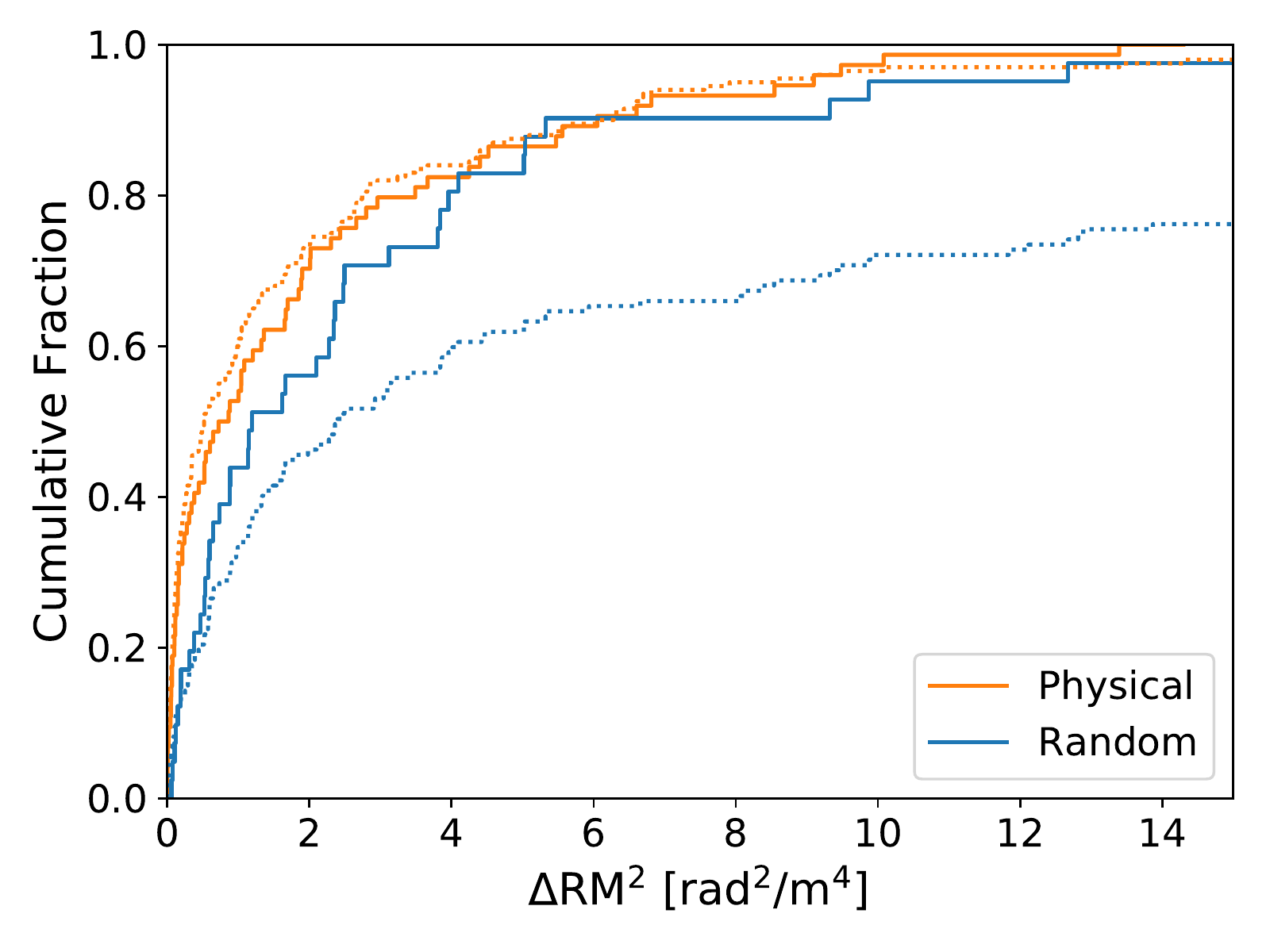}
\caption{ Empirical cumulative distribution functions (ECDFs) of the squared difference in 
RM, \dRMsq~in units of\radsq, between pairs of radio sources. 
The dashed blue and orange lines correspond to 
all the data for physical and random pairs, respectively, while the solid blue and orange lines show only 
the corresponding data for physical and random pairs in the overlapping region of angular separation 
from 2 to 10~arcmin. }\label{fig:ecdf}
\end{figure}

The root-mean-square of the RM difference for all RPs, \rms$_{,{\rm RP}} = 6.0\pm0.5$\rad~while 
 \rms$_{,{\rm PP}} = 1.4\pm0.1$\rad~(in all cases we quote the rms with the error variance subtracted, and 
 the associated uncertainty is the standard error in the mean). 
 Kolmogorov-Smirnov (KS) and Anderson-Darling (AD) 
tests indicate that the difference between RPs and PPs is significant at $>5\sigma$ (with p-values of $\sim$$10^{-7}$ 
and $\sim$$10^{-4}$ respectively). 
The empirical cumulative distribution function (ECDF) of \dRMsq~is shown in Fig.~\ref{fig:ecdf}, with all 
PPs and RPs shown with dotted lines. 
This difference is dominated by the general trend of larger \dRM~variations at larger angular separations (Fig.~\ref{fig:dRMsq_sep}), 
as is expected if the Milky Way ISM is a significant contributing factor to the RM variance on these angular scales \citep[e.g.][]{stil2011}. 

\begin{table}
\centering
\caption{Summary of results in the RM difference (\dRM) between pairs.}
\label{tab:basic}
\begin{tabular}{lccccc}
\hline
Classification & $N$ & $\Delta{\rm RM_{rms}}$ & $\Delta{\rm RM_{rms*}}$ & $|\Delta{\rm RM}|_{\rm median}$   \\
 & & rad m$^{-2}$  & rad m$^{-2}$  & rad m$^{-2}$  \\
\hline
Random pairs (RP)         & 148     &  6.0$\pm$0.5     &  5.5$\pm$0.4   &   1.5$\pm$1.4         \\
Physical pairs (PP)         &  201    &  1.4$\pm$0.1      &  1.4$\pm$0.1  &   0.7$\pm$0.6          \\
RP: 2 to 10 arcmin  &   42     &  5.1$\pm$0.8  &  1.8$\pm$0.3 &    1.2$\pm$0.6       \\
PP: 2 to 10 arcmin  &   75     &  1.4$\pm$0.2   & 1.4$\pm$0.2 &    0.9$\pm$0.6        \\
\hline
\end{tabular} 
\scriptsize{$\Delta{\rm RM_{rms*}}$: Trimmed rms, with one outlier from the random pairs removed. }
\end{table}

If we restrict the comparison only to the region of significant overlap in angular separation (i.e.~2 to 10~arcmin), then we have 
\rms$_{,{\rm RP}} = 5.1\pm0.8$\rad~and \rms$_{,{\rm PP}} = 1.4\pm0.1$\rad. However, 
the rms for the RPs is strongly affected by 
one outlier, with the highest value of \rms~in the sample 
of $\sim$954\radsq~(highlighted by a box in Fig~\ref{fig:dRMsq_sep}). 
Therefore, we introduce the ``trimmed rms'' ($\Delta{\rm RM_{rms*}}$) with this outlier removed. 
This reduces the rms of the RPs to \rms$_{,{\rm RP}} = 1.8\pm0.3$\rad, giving a difference of $0.4\pm0.3$\rad~between 
the RPs and PPs between 2 and 10~arcmin. 
The difference in the median values of $|\Delta{\rm RM}|$ for RPs and PPs in the overlap region 
is $0.3\pm0.8$\rad. For the uncertainties in the median $|\Delta{\rm RM}|$, we use the median 
absolute deviation (i.e.~half the interquartile range). 
These results are summarised in Table~\ref{tab:basic}.  
The ECDFs of \dRMsq~for only those RPs and PPs in the overlapping angular separation region of 2 to 10~arcmin 
are shown with solid lines in Fig.~\ref{fig:ecdf}.
KS and AD tests indicate that the RPs and PPs in the overlap region are not significantly different 
for these sample sizes (p-values of 0.17 and 0.06, respectively). 
The exclusion of the outlier does not significantly affect the KS or AD test results (p-values of 0.20 and 0.08). 
Therefore, based on the difference in the trimmed rms ($\Delta{\rm RM_{rms*}}$) values of $0.4\pm0.3$\rad~and 
the difference in the median values of $|\Delta{\rm RM}|$ of $0.3\pm0.8$\rad, we consider a conservative  
upper limit on the excess Faraday rotation contribution between RPs to be 
1.9\rad~(i.e.~the median difference plus twice the uncertainty).  
We use this upper limit to derive a model limit on extragalactic magnetic fields in Section~\ref{sec:basicsims}. 

\subsection{Trends in \dRM~as a function of angular separation}\label{sec:sfrm}

We also fit a power-law function to the RPs and PPs data, $(\Delta{\rm RM})^2=k\,\Delta\theta^\gamma$, 
where $k$ is a constant with units of rad$^2$~m$^{-4}$~arcmin$^{-\gamma}$. We find $k_{\rm RP}=0.6\pm1.0$, 
$\gamma_{\rm RP}=0.6\pm0.4$ and $k_{\rm PP}=0.25\pm0.04$, $\gamma_{\rm PP}=0.8\pm0.2$. 
These fits are shown in Fig.~\ref{fig:dRMsq_sep}. 
The difference in \dRMsq~between the power-law fits at the average separation in the overlapping region (i.e.~6') is $\sim$0.8\radsq. 
Attempts at fitting only the data in the overlapping region were poorly constrained, so we do not include them here. 

\begin{figure}
\includegraphics[width=8.2cm,clip=true,trim=0.4cm 0.6cm 0.0cm 0.0cm]{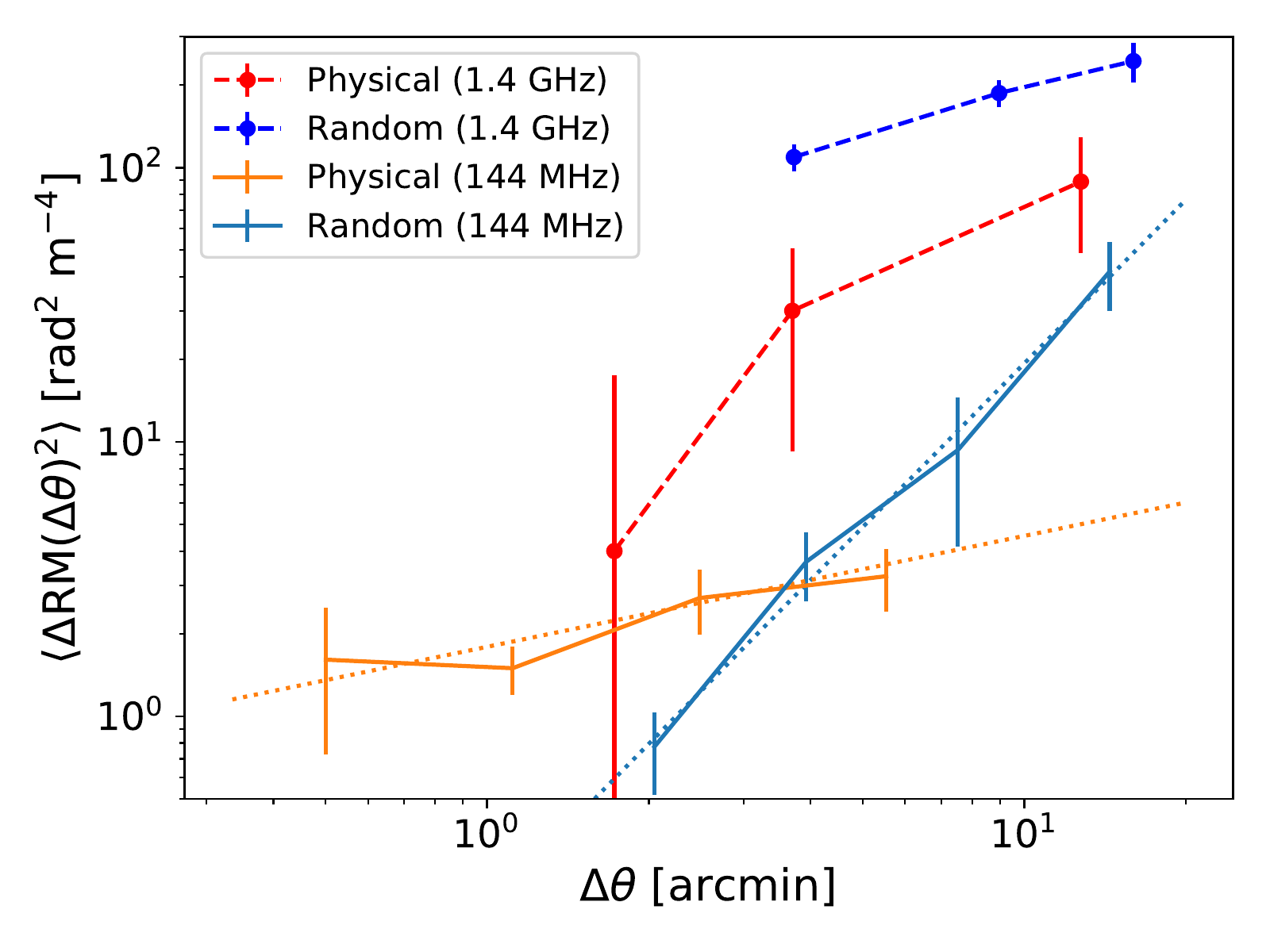}
\caption{Plot of the RM structure functions (i.e.~mean bins of \dRMsq~as a function of the 
pair angular separation, $\Delta\theta$) with the noise power from measurement errors subtracted, for PPs (orange) and RPs (blue) 
at 144~MHz. The orange and blue dotted lines show the power-law fits to the PPs and RPs, respectively. 
The RPs bin at the smallest angular separation has only 7 data points, and may be unreliable. 
For comparison, we also include the structure function results from the 1.4~GHz data of \citet[][]{vernstrom2019}  
for PPs (red circles) and RPs (dark blue circles), connected by dashed lines. 
}\label{fig:sfrm}
\end{figure}

To more directly compare with the results of V19, we calculate the mean of \dRMsq~as a function of 
the angular separation, i.e.~the RM structure function (SF), with
\begin{equation}
\langle \Delta{\rm RM(\Delta\theta)}^2 \rangle = \frac{1}{N}\sum_i [{\rm RM}_1(\theta)-{\rm RM}_2(\theta+\Delta\theta)]_i^2 \, .
\end{equation}
This is done separately for the RPs and PPs, and they are plotted in Fig.~\ref{fig:sfrm} in addition to the V19 RM structure functions. 
By fitting power-laws to these data in a similar manner to above, with 
$\langle (\Delta{\rm RM})^2 \rangle=k_{\rm SF}\,\Delta\theta^{\gamma_{\rm SF}}$, 
we find $k_{\rm SF,RP}=0.2\pm0.1$, $\gamma_{\rm SF,RP}=1.9\pm0.2$ (with the outlier removed) 
and $k_{\rm SF,PP}=1.8\pm0.3$, $\gamma_{\rm SF,PP}=0.4\pm0.1$. 
These fits are shown by dotted lines in Fig.~\ref{fig:sfrm}. 
The values of $\gamma_{\rm SF}$ are in stark contrast with those found in 
V19 ($\gamma_{\rm SF,RP,NVSS}=0.6\pm0.1$, $\gamma_{\rm SF,PP,NVSS}=0.8\pm0.2$) 
with the RPs slope being much steeper than in V19 and the PPs slope being much flatter. 
Also notable is that the overall amplitude is smaller in both cases compared to V19 
($k_{\rm SF,RP,NVSS}=50\pm30$, $k_{\rm SF,PP,NVSS}=11\pm15$).  
These differences have important implications for the nature of the extragalactic Faraday rotating material 
and are addressed in the next section. 

\subsection{Comparison with RM data at 1.4~GHz}\label{sec:v19compare}
We find that 203 out of the 698 source components in this study ($\sim$29\%) have corresponding RM values at 1.4~GHz 
in the \cite{taylor2009} catalog. 
The vast majority, 91\% (97\%), of the corresponding RM values are consistent within $3\sigma$ $(5\sigma$) of the combined RM error.  
The LOFAR sources that are not in the NVSS RM catalog are on average $\sim$3 times fainter in total intensity at 144~MHz. 
This means that the majority of the LOFAR polarized sources are too faint to be detected in the NVSS, while the 
majority of the NVSS sources are depolarized at LOFAR frequencies \citep[e.g.][]{osullivan2018b}. 
For those pairs that have counterparts in the V19 catalog, we find 
$\Delta{\rm RM_{rms,RP,NVSS}}\sim18$\rad~and $\Delta{\rm RM_{rms,PP,NVSS}}\sim5$\rad, 
which is consistent with the results presented in V19. 
However, for the exact same sources we find $\Delta{\rm RM_{rms,RP,LOFAR}}\sim5$\rad~and $\Delta{\rm RM_{rms,PP,LOFAR}}\sim2$\rad.  
Furthermore, the median degrees of polarization for these RPs and PPs at 1.4~GHz are $\sim$7\% and $\sim$11\%, respectively, 
while at 144~MHz they are significantly lower at $\sim$1.5\% and $\sim$3\%, respectively. 

This large difference in both the RM variance and degree of polarization of the same sources at 1.4~GHz and 144~MHz is 
most likely related to the broader range of Faraday depths that are probed local to the sources at 1.4~GHz. 
A plausible scenario is that the LOFAR observations are only sensitive to the low RM variance regions 
of these sources, and that the differences between the V19 results and those presented here are 
due to the RM properties of the local source environment. 
The difference in angular resolution between these studies (a factor of 3) may also play a  
role and a more detailed investigation is needed, including high angular resolution observations at 1.4~GHz. 

The above hypothesis is supported by comparison of the RM structure functions at 144~MHz and 
1.4~GHz (analysed in Section~\ref{sec:sfrm} and shown in Fig.~\ref{fig:sfrm}). The difference between 
the RP data at 1.4 GHz and 144~MHz is $\sim$10\rad, which is comparable to the total extragalactic 
RM variance estimated by V19, \citet{schnitzeler2010} and \citet{oppermann2015}. 
This likely reflects the typical contribution to the RM variance at 1.4 GHz provided by the 
magnetized intragroup/intracluster medium surrounding radio galaxies. In this case, the Faraday depolarization 
caused by these environments is sufficient to depolarize the majority of sources  
below the detection threshold at 144~MHz, leaving only the low RM variance regions of some of 
these sources detectable with LOFAR. 

The steeper slope of the RM structure function for RPs (compared to V19, see Fig.~\ref{fig:sfrm}) 
may more cleanly reflect the RM variance from the Milky Way on these angular scales, if the extragalactic 
RM variance contribution is indeed much lower for the LOFAR data (more data for the PPs at large angular 
separations are needed to test this hypothesis). 
The flattening of the slope of the RM structure function towards smaller angular scales 
(as probed by the PPs), may reflect a growing contribution of the extragalactic RM variance (relative to 
the Milky Way), with $\langle$\dRMsq$\rangle \lesssim1.6$\radsq~on the smallest angular separations (Fig.~\ref{fig:sfrm}). 
We use this limit as a constraint for cosmological MHD simulations in Section~\ref{sec:MHDsims}. 

\begin{figure}
\includegraphics[width=8.2cm,clip=true,trim=0.4cm 0.6cm 0.0cm 0.0cm]{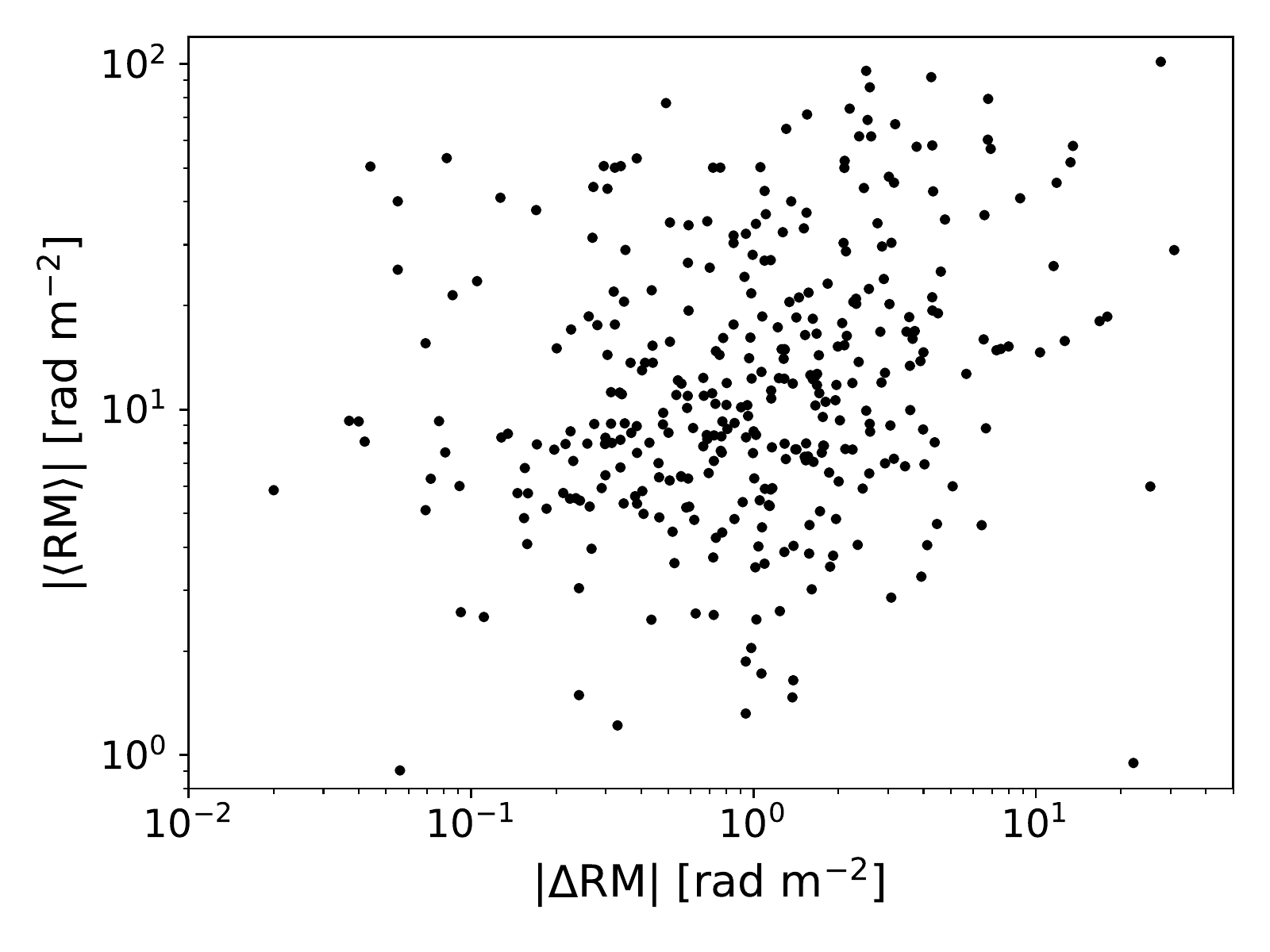}
\caption{ Plot of the average RM, $|\langle{\rm RM}\rangle|$, versus 
the RM difference, $|\Delta{\rm RM}|$, for each pair. A Spearman rank 
test indicates they are weakly correlated, with a correlation coefficient 
of 0.2, at a significant level (p-value $\sim10^{-5}$). 
 }\label{fig:avgRM}
\end{figure}

If the Milky Way dominates the RM variance, then we might expect the average RM of each pair 
to be correlated with $\Delta{\rm RM}$, because the average RM is known to be dominated by 
the Milky Way \citep[e.g.][]{oppermann2012}. In Fig.~\ref{fig:avgRM} we plot the absolute value 
of the average RM, $|\langle{\rm RM}\rangle|$, versus the absolute value of the RM difference, 
$|\Delta{\rm RM}|$, for each pair. A Spearman rank test indicates these quantities are weakly 
correlated (correlation coefficient of 0.23) with a significance of $\sim4.4\sigma$ (p-value $\sim10^{-5}$). 
The correlation for PPs is slightly stronger (0.26, p-value: $\sim10^{-4}$) than for RPs 
(0.20, p-value: $\sim10^{-2}$). 
This indicates that, as expected, the Milky Way contributes to the RM variance even on these 
small angular scales. However, it remains unclear what the exact contribution is relative to the 
extragalactic RM variance. A much higher surface density of polarized sources on the sky is 
required to accurately estimate the Milky Way RM contribution for this dataset. 

\section{Simulations}\label{sec:sims}

\subsection{Monte Carlo Modelling of Extragalactic RM Pairs}\label{sec:basicsims}

In order to understand the implications for the properties of intergalactic magnetic fields 
based on the results in Section~\ref{sec:results}, we develop some basic simulations of 
extragalactic Faraday rotation.  
We use a model of an inhomogeneous evolving universe, with an initial cosmological 
magnetic field, the strength of which is scaled with the local density variations. 
This model allows us to calculate the RM along cosmic sight lines to PPs and 
RPs for different angular and redshift separations. 
We investigate a wide range of initial magnetic field strengths and 
correlation lengths, which are then constrained by the RM observations. 
In Section~\ref{sec:limitations}, we comment on some of the limitations of this model and 
compare this approach with simpler models in Section~\ref{sec:homo}. 

Following \cite{blasi1999} and \cite{pshirkov2016}, we model the electron number density 
along cosmic lines of sight as $n_e(z)=n_e(0)(1+\delta_e)(1+z)^3$, 
with $n_e(0)=1.8\times10^{-7}$~cm$^{-3}$ and $\delta_e$ being the electron overdensity. 
We draw the electron overdensity $\delta_e$ from a log-normal distribution, with $\delta_e$ varying on scales 
of the Jeans length, $\lambda_J(z)\sim2.3(1+z)^{-1.5}$~Mpc. The log-normal distribution is given by 
\begin{equation}\label{eqn:lognormal}
P(\delta_e) = \frac{1}{\sqrt{2\pi}\sigma_e(1+\delta_e)} \exp\left\{-\frac{\left[\ln(1+\delta_e)-\mu_e(z)\right]^2}{2\sigma_e^2}\right\}\, , 
\end{equation}
where the mean ($\mu_e$) and standard deviation ($\sigma_e$) of the logarithm of the $\delta_e$ distribution are 
constrained from observations of the Lyman-$\alpha$ forest \citep[e.g.][]{bidavidsen1997}, with 
$\sigma_e(z)=0.08+5.37(1+z)^{-1}-4.21(1+z)^{-2}+1.44(1+z)^{-3}$ and $\mu_e(z)=-\sigma_e(z)^2/2$. 
For close pairs of sources we also need to include a prescription for the correlation of the densities along 
adjacent lines of sight. For this we use the two-point galaxy correlation function 
$\xi(r,z)=(r/r_0)^{-\gamma}(1+z)^{-(2+\gamma)}$, including its expected redshift evolution, 
with $\gamma\sim1.8$ and $r_0\sim5h^{-1}$~Mpc \citep[e.g.][]{vandenbosch2010}. 
We consider this correlation function valid for scales between $0.2h^{-1}$ to $30h^{-1}$~Mpc. 
Practically, we implement the correlated draws using a bi-variate Gaussian distribution before 
taking the exponential, where the off-diagonal terms of the covariance matrix are given by 
$\xi_G={\rm ln}(1+\xi)$ \citep[e.g.][]{coles1991,chuang2015,baratta2019}.
For separations between adjacent cells larger than $30h^{-1}$~Mpc, we draw from an 
uncorrelated log-normal distribution, while the same density is assigned for cell 
separations less than $0.2h^{-1}$~Mpc. 

We model the magnetic field strength as a scaled function of the density and redshift following $B(z)=B_0[n_e(z)/n_e(0)]^{2/3}$, 
which is a reasonable expectation in the case of isotropic gas compression \citep[e.g.][]{locatelli2018}.  
In this case, we have $B_0$ as the co-moving cosmological magnetic field strength in nG.  
The correlation length of the magnetic field ($l_B$) is set in fractions of the Jeans length, with random orientations 
assigned at each step by multiplying the amplitude of the field by a number drawn from a uniform [$-1$, $1$] distribution. 
The RM values were obtained by summing the RM contributions over all cells along an individual line of sight (from the source to us), 
while accounting for the RM redshift dilution of $(1+z)^2$ in each cell. 

Since we do not know the redshift distribution of our sample, we randomly draw sources from a
log-normal redshift distribution ($\mu_z=-1$, $\sigma_z=1$), 
which has a median of $z\sim0.37$ and is consistent with the redshift distribution of polarized 
extragalactic radio sources \citep{vernstrom2019,hardcastle2019,osullivan2018b}. 
However, we limit the redshift to a maximum of 1 for the PPs and 4 for the RPs following the corresponding maximum 
redshifts found in V19. 
We note that the actual redshift distribution of the LOFAR data may be somewhat different because 
the sources are typically fainter in total intensity that those at 1.4~GHz (Section~\ref{sec:v19compare}).
Similarly, we do not know the projected linear size distribution of our physical pairs, but we can again reasonably 
model this as a log-normal distribution ($\mu_{\rm ls}=-1$, $\sigma_{\rm ls}=1$) 
in Mpc units based on the projected linear size distributions for LOFAR radio galaxies in \cite{hardcastle2019}. 
For an angular size ranging from 2 to 10~arcmin, this gives a maximum linear size of $\sim$5~Mpc 
and a minimum of $\sim$24~kpc, which is consistent with the range of linear sizes of 
LOFAR polarized sources \citep{osullivan2018b}. 
In general, this aspect of the analysis can be substantially improved when the 
redshifts of the LOFAR polarized sources become available in the upcoming LoTSS DR2 value-added catalog.

For the PPs, we first draw the redshift of the radio galaxy, then the linear size, and compute the 
separation, $\theta$, between the pair using the angular diameter distance. 
We construct two sight lines to the radio galaxy, only allowing the range $2<\theta<10\,{\rm arcmin}$, to obtain the RM for each sight 
line, before calculating the RM difference (in a random manner). For the RPs, we draw a redshift for each radio galaxy, 
with a fixed $\theta$ of 6~arcmin (i.e.~the mean observational separation for the RPs in the overlap region), before 
calculating the RM difference. 
To create distributions of \dRM, we calculate the RM difference for 10,000 draws each 
for both RPs and PPs. This produced smooth distributions of \dRM~from which we could obtain reliable statistics. 

We then generate \dRM~distributions for RPs and PPs for a grid of $B_0$ and $l_B$ values. 
The simulations are run for a grid of $0.1 \leq B_0 \leq 10$~nG and $0.1 \leq l_B/\lambda_J \leq 10$, 
both with 10 even steps in log space. To extend the grid to large values of $l_B$, 
we also produce \dRM~distributions for $10 \leq l_B/\lambda_J \leq 1000$ 
in 5 even steps in log space (i.e.~for a total of 150 grid points). We employ a parallelized version of the code 
(using the joblib python library\footnote{https://joblib.readthedocs.io/en/latest/parallel.html}), 
which takes approximately 3 weeks to run on 24 cores. 
The median of $|\Delta{\rm RM}|$ was chosen as the most robust statistic for comparison with the 
observational data (see Table~\ref{tab:basic}). This is because the mean $(\Delta{\rm RM})^2$ values 
from the model are strongly affected by outliers, due to the lognormal density distribution \citep[e.g.][]{blasi1999}. 

\begin{figure}
\includegraphics[width=8.2cm,clip=true,trim=0.0cm 0.0cm 0.0cm 0.0cm]{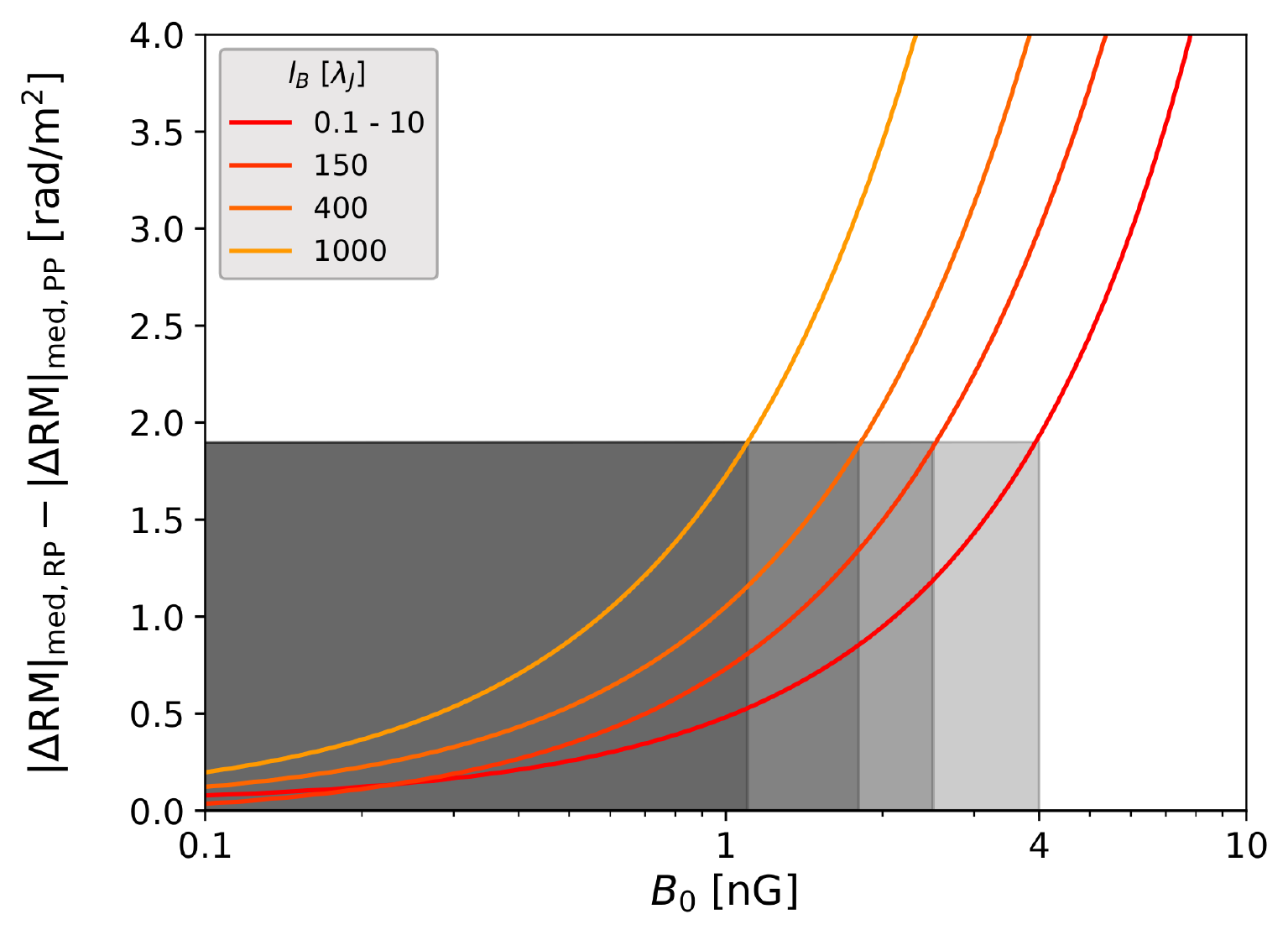}
\caption{ Plot of $|\Delta{\rm RM}|_{\rm median,RP} - |\Delta{\rm RM}|_{\rm median,PP}$ (in \rad)~versus the input initial 
cosmological magnetic field strength ($B_0$ in nG), provided by the inhomogeneous universe model, 
described in Section~\ref{sec:basicsims}. 
The lines trace the growth in the difference in the Faraday rotation between RPs and PPs for increasing 
values of $B_0$. The shaded regions outline the upper limits on $B_0$ for various magnetic field 
correlation lengths ($l_B$ in units of the Jeans length, $\lambda_J$), provided by the points at which 
the upper limit of 1.9\rad~(derived in Section~\ref{sec:results}) intersects with the lines. 
The line furthest to the right defines the upper limit of $B_0\lesssim4$~nG on Mpc scales. 
}\label{fig:grid}
\end{figure}

Figure~\ref{fig:grid} shows the value of $|\Delta{\rm RM}|_{\rm median,RP} - |\Delta{\rm RM}|_{\rm median,PP}$ 
for variations in $B_0$ and $l_B$.  
The shaded regions outline the allowed values of $B_0$ based on the limit of 1.9\rad~from Section~\ref{sec:results}.  
The lines show the dependence of the excess Faraday rotation contribution to RPs on $B_0$, for correlation 
lengths in the range $0.1 \leq l_B/\lambda_J \leq 1000$. Only small variations are present in the model output 
for $0.1 \leq l_B/\lambda_J \leq 10$ so we just show one line for the average dependence. 
This places a conservative limit on the co-moving cosmological magnetic field of $B_0<4$~nG 
for correlation lengths on Mpc scales (with smaller $B_0$ limits for larger $l_B$). This limit should not be confused 
with the magnetic field strength in overdensities, which can be an order of magnitude larger due to the scaling 
with density in this model. 

\subsubsection{Limitations of the Monte Carlo model}\label{sec:limitations} 
While the above model is a good approximation for spherical overdensities and underdense 
regions like voids, a major limitation of this approach is that it does not accurately  
describe the density variations expected in cosmic sheets and filaments. If the contribution of 
magnetic fields in sheets and filaments to the observed RM is significant, then it is plausible 
that the magnetic field limits from this model are overestimated.
Further limitations of the model are the assumption of how the magnetic field scales 
with the electron density, as well as the existence of a single correlation 
scale of magnetic fields along the line of sight.

Although limited, we consider this model an advance on models with a homogenous 
electron density distribution and magnetic field strength along the line of sight. The inhomogeneous 
model can be developed further by incorporating more realistic electron density distributions 
that better represent cosmic sheets and filaments, in addition to better modelling of the physical 
properties of the sources in the sample after obtaining their redshifts. 

\subsubsection{Alternative approaches to magnetic field limits}\label{sec:homo}

The simplest model one can adopt is of a homogeneous universe, where the excess 
\rms~of $<1.9$\rad~comes from a uniform IGM along the line of sight between the RPs. 
In this case, we take the median redshift of the RPs to be 0.4 in order to estimate the average 
electron density of $n_e = n_e(0)(1+z)^3 \sim 5\times10^{-7}$~cm$^{-3}$, with an rms magnetic 
field in the IGM ($B_{\rm IGM,rms}$) having a coherence length ($l$) of 1~Mpc. We take the typical 
distance between the RPs to be $L\sim1$~Gpc (using the median $\Delta{z}$ of $\sim$0.4 between RPs found in V19). 
This leads to a limit of $B_{\rm IGM,rms}<260$~nG, for 
\begin{equation}
\sigma_{\rm RM,ex}< 1.9 
\left(\frac{B_{\rm IGM,rms}}{260{\rm~nG}} \right) 
\left(\frac{n_e}{5\times10^{-7}{\rm~cm}^{-3}} \right) 
\left(\frac{l}{1{\rm~Mpc}} 
\frac{L}{1{\rm~Gpc}} \right)^{1/2}  
{\rm~rad~m}^{-2}.
\end{equation}
However, as we expect the RM signal to be dominated by overdense regions along the line of sight \citep[e.g.][]{akahoriryu2011}, 
we consider the limits from the inhomogeneous model more appropriate, even though our inhomogeneous model does not 
accurately describe the filamentary structure of the cosmic web as seen in cosmological simulations. 

Alternatively, one could assume the extragalactic RM variance between RPs is completely dominated 
by cosmic sheets and filaments (with an insignificant contribution from voids). In this case a limit on the 
rms magnetic field strength in the sheets and filaments ($B_{\rm filament}$) can be estimated. 
Using the same coherence 
length and path length as above, and assuming that 25\% of the line of sight between RPs (i.e.~$f L \sim 0.25$~Gpc) 
is intersected by sheets or filaments \citep[e.g.][]{cautun2014}, we find that an rms 
magnetic field strength of $\sim$26~nG and an average electron density of 10$^{-5}$~cm$^{-3}$, 
could provide 
\begin{equation}
\sigma_{\rm RM,ex}< 1.9 
\left(\frac{B_{\rm filament}}{26{\rm~nG}} \right) 
\left(\frac{n_{e,{\rm filament}}}{10^{-5}{\rm~cm}^{-3}} \right) 
\left(\frac{l}{1{\rm~Mpc}} 
\frac{f L}{1{\rm~Gpc}} \right)^{1/2}  
{\rm~rad~m}^{-2}.
\end{equation}
If we further assume that the magnetic field in the filaments scales from an initial cosmological field 
as $(n_{e,{\rm filament}}/n_e(0))^{2/3}$, then the initial 
field would be $\sim$2~nG (within a factor of 2 of our limit of 4~nG in Section~\ref{sec:basicsims}). 

\subsection{Comparison with cosmological simulations}\label{sec:MHDsims}

For constraints based on a more realistic model of the universe, we use the results of recent magneto-hydrodynamical (MHD) simulations
considering several different scenarios for the origin and amplification of extragalactic magnetic fields \citep[][]{vazza2017}. 
The comparison that we focus on in this case is with the LOFAR RM structure function for the PPs 
on small angular scales (c.f.~Fig.~\ref{fig:sfrm} for $\Delta\theta<1$~arcmin). 
We consider this the most relevant 
constraint because the extragalactic contribution to the structure function may begin to dominate at small 
angular scales since the contribution of the Milky Way ISM is expected to 
decline steeply with decreasing angular scale \citep[e.g.][]{akahori2013,akahori2014}. 
In contrast to the model described in Section~\ref{sec:basicsims}, here the 3-dimensional distribution 
of magnetic fields and electron density in the cosmic volume are self-consistently produced 
by the MHD simulation, depending on different assumed scenarios for magnetogenesis. 
Here we focus on three of the most realistic scenarios, within a larger survey of 25 models presented in \citet{vazza2017}. 
A detailed survey of all models allowed by the LOFAR data is beyond the scope of the current work, 
but will become more feasible when we know the redshift distribution of the LOFAR sources. 

The three different scenarios were simulated in a comoving $(85\ \mathrm{Mpc})^3$ volume with $1024^3$ cells, 
using the ENZO code \citep[][]{vazza2017}. The different prescriptions for the injection and evolution of magnetic fields were: 
a) a primordial, uniform, volume-filling, comoving magnetic field of $B_0=0.5\ \mathrm{nG}$ at the beginning of the simulation; 
b) a primordial model starting from the much lower level of $B_0 = 10^{-9}\ \mathrm{nG}$ but including a run-time modelling 
of dynamo amplification of the magnetic field; 
c) an ``astrophysical'' run in which the magnetic field is injected solely by feedback events from star forming regions and/or active galactic nuclei. 
For scenario a), a residual level of magnetization ($\sim$1 to 10~nG) is present everywhere in the cosmic volume. 
However, in scenarios b) and c) the average magnetization is a steeper function of density. 
Particularly in the astrophysical scenario, very little magnetic fields are present outside of the virial volume of matter halos, due to 
the strong association between sources of magnetization and the halos.  For more details we refer the reader to 
\citet{vazza2017} and \citet{gheller2019}. 

In order to construct synthetic RM structure functions for the PPs for each magnetogenesis scenario, simulated maps of 
Faraday rotation for a $4^\circ$ field of view were created, before obtaining deep lightcones by stacking different 
snapshots along the line of sight. 
We note that the generation of synthetic RM structure functions for RPs was beyond the scope of the 
current work, but will be investigated in a future publication. 
In detail, using different snapshots of the above runs, we integrated along lightcones up to $z=0.5$, 
and simulated $(\Delta{\rm RM})^2$ for PPs, by placing pairs of sources at 
regular intervals of $85 \rm ~Mpc$ (co-moving) along the line of sight (i.e.~at the end of each of the 
co-moving volumes used to produce the stacked sequence of Faraday rotation). 
We first randomly drew $1500$ sources, with $|\rm RM| \geq 0.03$\rad, for $22$ evenly spaced redshift bins. 
We then computed the $\Delta {\rm RM}(\Delta \theta)^2$ statistics at each redshift, and finally produced the 
observable total distribution of $\Delta {\rm RM}(\Delta \theta)^2$ by weighting each structure function by the distribution 
function of source redshifts approximately derived from V19. 

\begin{figure}
\includegraphics[width=9cm,clip=true,trim=0.0cm 0.0cm 0.0cm 0.0cm]{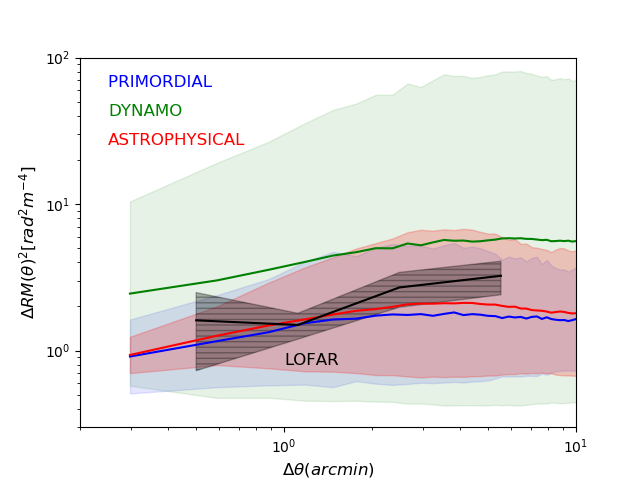}
\caption{Simulated distribution of $\Delta {\rm RM}(\Delta \theta)^2$ as a function of angular separation 
for three numerical models, as described in Section~\ref{sec:MHDsims}, compared with LOFAR data. 
The solid lines show the mean values and the 
shaded region shows the $1\sigma$ dispersion. The dark shaded region outlining the LOFAR data is 
identical to that shown in Fig.~\ref{fig:sfrm} for the PPs.  
The blue line gives the prediction for a uniform primordial model of $B_0=0.5$ nG (co-moving). 
The variance around each model is due to the redshift distribution of sources. }
\label{fig:MHDsims}
\end{figure}

Figure \ref{fig:MHDsims} shows the simulated distribution of $\Delta {\rm RM}(\Delta \theta)^2$ as a function of 
angular separation for the three models, enabling a direct comparison with the LOFAR data (dark shaded region). 
Note that the RM variance from the Milky Way is not included in the models (so the model trends 
are not expected to exceed the LOFAR curve). The dynamo amplification model (green) is quite far 
from our LOFAR observations. Based on the typical range of magnetic field measured within filaments 
in these runs \citep[][fig.~6]{gheller2019}, this suggests a limit of $\lesssim 10$ to 100~nG on the average 
magnetisation of filaments crossed by the polarised emission observed with LOFAR. On the other hand, 
the astrophysical scenario (red line) and the uniform primordial model (blue line) give a more reasonable 
agreement with the LOFAR data, which follows from the fact that in this case the magnetic fields in filaments 
are far less volume filling, with a broad distribution of values centred around 1 nG. This in turn suggests that 
$B_0 \approx 0.5 \rm ~nG$ is the upper limit on primordial magnetic fields that can be derived from the LOFAR data. 
Conversely, no rescaling can reconcile the simulated statistics of $\Delta {\rm RM}(\Delta \theta)^2$ in the dynamo case, 
because the observed scatter in the LOFAR observations is more than one order of magnitude smaller than what is produced 
by the large fluctuations that are present across the distribution of filaments in the dynamo scenario \citep[e.g.][]{ryu2008}. 

We note that the constant spatial resolution of the cosmological simulations (83 kpc/cell) means that scales 
below 1~arcmin are not resolved by the simulation for the $z \leq 0.07$ portion of the lightcone. 
This likely makes our simulated distribution of (\dRM)$^2$ for $\Delta\theta\leq 1$~arcmin 
a lower limit of the true distribution that can be expected for each model.  
It is also important to note that the simulated pairs were not placed at the physical location of the AGN outflows.  
In general, this was done to avoid a strong contribution from the ambient AGN medium to all sources, 
which would have the effect of increasing the RM variance for a fixed primordial field strength. 
To remain consistent with the LOFAR data, this would then have pushed the magnetic field 
limits even lower and made assessing the dynamo amplification scenario in filaments difficult. 
Our approach is also potentially more consistent with the data because we do not expect to detect 
polarized emission from LOFAR sources embedded in dense magnetoionic media \citep[e.g.][]{stuardi2020}. 
This means the dynamo scenario is disfavoured solely in cosmic filaments, and not in denser regions like in the intracluster medium. 
The ability to model magnetic field fluctuations on small-scales would need to be 
added to the simulations before a more realistic distribution of radio galaxy environments could be included. 
Our hypothesis for the difference between the V19 and LOFAR results (i.e.~the difference being due 
to the ambient radio galaxy medium) could be tested with such an implementation. 

\section{Discussion}\label{sec:discussion}
The goal of this work is to isolate the extragalactic RM variance from the other contributions along the line of sight (e.g.~Eqn.~\ref{eqn:rmvariance}). 
The RM variance introduced by the ionosphere ($\sigma_{\rm RM,ion}^2$) was accounted for by restricting the RM difference of 
close pairs of radio sources, (\dRM)$^2$, to come from the same observational pointing, in addition to the standard correction of the 
time-variable ionosphere RM as described in Section~\ref{sec:data}. 
The RM variance contributed by measurement errors ($\sigma_{\rm RM,err}^2$) 
was subtracted from the quoted rms values, although the effect of this is minor due to the small LOFAR RM errors ($\sim0.03$\rad). 
We then split the (\dRM)$^2$ sample into pairs from the same physical source (physical pairs; PPs) and non-physical, random pairs 
on the sky (random pairs; RPs). The comparison of these two samples can then be used to account for the 
Milky Way contribution ($\sigma_{\rm RM,MW}^2$) in a statistical sense. 
In principle, this leaves only the contribution from extragalactic Faraday rotation ($\sigma_{\rm RM,ex}^2$).  
By analysing the difference in (\dRM)$^2$ between RPs and PPs in 
Section~\ref{sec:results}, we limit the extragalactic RM contribution between the LOFAR RPs to $<1.9$\rad~($\sim$95\% confidence). 

\subsection{The Faraday medium local to radio sources}

For the discussion, we split the extragalactic RM variance into contributions local to the 
source ($\sigma_{\rm RM,local}^2$) and from the intergalactic medium in the more distant foreground ($\sigma_{\rm RM,IGMF}^2$). 
At 1.4~GHz, V19 found an rms difference of $\sim$5 to 10\rad~between RPs and PPs. 
This is similar to $\sigma_{\rm RM,ex}\sim7$\rad~estimated by \citet{schnitzeler2010} and \citet{oppermann2015} at 1.4~GHz. 
However, our result at 144~MHz (i.e.~$\sigma_{\rm RM,ex}<1.9$\rad) appears to be in conflict with the 1.4~GHz results, as 
one does not expect a strong frequency-dependent $\sigma_{\rm RM,IGMF}$. 
We investigated this further in Section~\ref{sec:v19compare}, where we found that a)~the majority of the 144~MHz polarized sources were 
not detected in the NVSS at 1.4~GHz (because they are too faint, as LoTSS is $\sim$10 times more 
sensitive for steep spectrum sources), 
b)~most polarized sources at 1.4~GHz are not detected at 144~MHz (due to Faraday depolarization), and importantly
c)~the polarized sources in common have a smaller \rms~(and degree of polarization) at 144~MHz. 
This indicates that the larger rms difference of $\sim$5 to 10\rad~found in V19 between RPs and PPs is 
due to RM variance in the magnetised environment local to the sources (i.e.~from $\sigma_{\rm RM,local}^2$). 

For example, for a polarized signal to be detected at 144~MHz, only small amounts of Faraday depolarization within the LOFAR 
synthesised beam are possible (e.g.~less than 0.4\rad~on scales $<20$~arcsec for the most common model of external Faraday dispersion, \cite{burn1966}, 
with $p(\lambda)\propto e^{-2\sigma_{\rm RM}^2\lambda^4}$). Alternatively, the polarized signal may originate from a compact emission 
region on sub-beam scales (e.g.~hotspots), and thus the inferred Faraday depolarization 
would not represent the RM variance on larger scales (i.e.~as would be relevant for physical pairs on scales $>100$~kpc). 

In any case, for radio sources in regions of dense magnetized gas, such as near the centre of groups and clusters of 
galaxies, there is likely too much Faraday depolarization for emission to be detected at 144~MHz. 
Furthermore, large asymmetries in the Faraday rotation properties of opposite lobes are often found 
in these rich environments due to, for example, the Laing-Garrington effect \citep{laing1988,garrington1988}, which 
would make the detection of polarized emission from physical pairs less likely compared to physically large radio sources 
that are closer to the plane of the sky \citep[e.g.][]{saripalli2009}. 
This is consistent with recent results that find the majority of polarized detections in LOFAR data are from hotspots of 
FRII radio galaxies that are not associated with galaxy clusters and have large physical sizes \citep{osullivan2018b,stuardi2020,mahatma2020}.  
Therefore, we expect that LOFAR polarized sources are typically located in regions of the Universe with low RM variance. 
This makes them ideal probes of the weak magnetization of the cosmic filaments and voids far from galaxy cluster environments. 

\subsection{Model limits on intergalactic magnetic fields}

In Section~\ref{sec:sims}, we take two approaches to deriving upper limits on the co-moving cosmological magnetic field strength. 
In one approach, we use a Monte Carlo model to generate distributions of \dRM~for RPs and PPs 
in a universe with an inhomogeneous matter distribution and with a magnetic field strength that scales with 
the density inhomogeneities (i.e.~$B\propto n_e^{2/3}$). 
The model allows us to explore a wide range of input co-moving cosmological magnetic field strengths ($0.1 \leq B_0 \leq 10$~nG) 
and correlations lengths ($0.1 \leq l_B/\lambda_J \leq 1000$). 
For this model, we find that the median $|$\dRM$|$ is the best statistical indicator due to the highly non-Gaussian 
\dRM~distribution.  
Using the observational constraint of the difference in the median $|$\dRM$|$ between RPs and PPs being $<1.9$\rad~provides 
an upper limit of $B_0<4$~nG for magnetic field correlation lengths in the range $0.1\leq l_B/\lambda_J  \leq 10$ 
(where $\lambda_J\sim2.3$~Mpc at $z=0$). 
This limit is comparable with upper limits on the primordial field from CMB measurements \citep{PLANCK2015},  
and almost 10 times lower than the upper limit of $\sim$37~nG derived in V19. 

In the second approach, we compare our observational results with cosmological MHD simulations \citep{vazza2017} in three 
different scenarios: a) a strong initial primordial field of $B_0=0.5$~nG, b) a primordial field of $B_0=10^{-9}$~nG with dynamo amplification, and 
c) magnetization only from AGN and galactic outflows. 
In this approach, the most useful constraint comes from the RM structure function on the smallest angular scales because 
this should have the smallest contribution from the RM variance of the Milky Way (which is not included in the models). 
In particular, synthetic RM structure functions for PPs were created from the simulations (as described in Section~\ref{sec:MHDsims}) 
and constrained by the data for angular separations less than 1~arcmin (i.e.~$\langle$\dRMsq$\rangle<1.6$\radsq). 
Both the scenario of magnetization by astrophysical processes (e.g.~AGN and galactic outflows) and the primordial 
case are consistent with the data, for an initial (spatially uniform) primordial seed field of $B_0\lesssim0.5$~nG. 
The dynamo amplification scenario is inconsistent with the data as it produces \dRM~fluctuations that are too large. 

These inferences can be considered preliminary, since one of the main limitations of the comparison between the models 
and the data is our lack of knowledge of the exact redshift distribution of the observed radio sources. 
For example, we do not know the true distribution of physical (and angular) separations for PPs as a function of redshift. 
Also, the simulated PPs are not placed at the location of AGN outflows (i.e.~no model contribution of  
$\sigma_{\rm RM,local}^2$), which leads to more conservative upper limits on the seed field and provides 
constraints that are more relevant to the dynamo amplification of field in filaments 
(rather than in more dense regions such as near galaxy clusters). 
Furthermore, we have not included a model for the RPs, mainly due to the lack of redshift information. 
For future work, in addition to more realistic models based on observed redshifts and environments, we plan to 
explore how the LOFAR data might also constrain the morphology of primordial magnetic fields, 
whose initial spectra are already constrained by PLANCK observations \citep{PLANCK2015}. 
In general, this highlights the potential of LOFAR data to realistically discriminate between competing  
magnetogenesis scenarios. 

\subsection{Upcoming advances}
Much more can be achieved in the near future with LOFAR. In particular, we expect the sample of pairs to potentially increase 
by an order of magnitude for the full LoTSS survey, helping to push well into the sub-nG regime for the study of cosmic magnetic fields. 
In the near term, host galaxy identifications and redshifts will be provided by the value-added data products in LoTSS DR2. We expect 
to get photometric or spectroscopic redshifts for $\sim$80\% of the polarized sources in our current sample \citep{osullivan2018b}. 
With the LOFAR-WEAVE survey \citep{smith2016}, we expect spectroscopic redshifts for all the polarized radio sources in LoTSS 
up to at least $z=1$. 
In combination with the high-fidelity 6~arcsec total intensity images provided by the LoTSS survey, these redshifts will 
enable precise linear size estimates of the sources, which will further enhance our ability to distinguish between 
magnetoionic material local to the source and that associated with cosmic filaments and voids. In addition, splitting the sample 
into redshift bins (in addition to $\Delta{z}$ bins for the random pairs) will allow investigations of the evolution of magnetic fields with cosmic time. 

In order to learn more about the properties of LOFAR polarized sources (and the IGMF), we will need to consider several other properties, 
such as the degree of polarization/depolarization, the total intensity spectral index, the radio source morphology, the environment, etc.    
Such investigations are important to allow a better understanding of the different astrophysical contributions to 
the total observed RM variance, to weight the RM variance of each sub-population in an appropriate manner \citep[e.g.][]{rudnick2019}, 
and to potentially remove blazars from the sample. 
This should be done in combination with other upcoming RM surveys at higher frequencies 
\citep[e.g.~POSSUM, VLASS;][]{possum,lacy2019}, 
which can probe cosmic magnetic fields in high density environments that are currently inaccessible for LOFAR. 
In the longer term, both the SKA-Low and SKA-Mid \citep[e.g.][and references therein]{braun2015} will be essential 
to further map out the frequency-dependent behaviour of the extragalactic RM variance in order to uncover the nature 
of magnetic fields in the cosmic web. 

\section{Conclusions}\label{sec:conclusions}

We have presented a Faraday rotation study of 349 close pairs of extragalactic radio sources with LOFAR, to investigate  
the properties of extragalactic magnetic fields. 
The data used are from the ongoing LOFAR Two-Metre Sky Survey \citep[LoTSS;][]{shimwell2019}, which is 
imaging the northern sky in continuum polarization from 120 to 168~MHz. The large bandwidth at such low 
frequencies provides exceptional RM precision, with typical errors of $\sim$0.03\rad, which are $\sim300$ times 
better than available for previous studies \citep[e.g.][]{vernstrom2019}. 

By considering the variance of the RM difference between physical pairs (e.g.~double-lobed radio galaxies) and 
non-physical, random pairs (i.e.~physically different sources with close projected separations on the sky), we 
statistically separate the extragalactic component of the RM variance from that due to the Milky Way.  
In the region of overlapping angular scales from 2 to 10~arcmin, we find a trimmed rms RM difference 
of $1.8\pm0.3$\rad~for 41 random pairs and $1.4\pm0.2$\rad~for 75 physical pairs, providing an 
estimate of $+0.4\pm0.3$\rad~for the excess Faraday rotation experienced by random pairs. 
A similar estimate of $+0.3\pm0.8$\rad~is found from an analysis of the median $|$\dRM$|$. 
A Kolmogorov-Smirnov test indicates that there is no significant difference between the \dRM~distributions 
of random and physical pairs in the region of overlapping angular scales. 
Using the difference in the median $|$\dRM$|$ values, we place an upper limit of 
$1.9$\rad~($\sim$95\% confidence) on the excess extragalactic Faraday rotation 
contribution to random pairs over physical pairs.

This result is in apparent conflict with estimates of the extragalactic variance of $\sim$5 to 10\rad~derived from 
observations at 1.4~GHz \citep{vernstrom2019}. There is no expectation of a frequency-dependent RM from 
magnetic fields in cosmic filaments and voids. Therefore, our results point to the contribution of magnetoionic 
material local to the radio source as the dominant extragalactic contribution at 1.4~GHz (e.g.~the magnetized IGM of galaxy groups and clusters). 
This means that sources in \citet{vernstrom2019} with large RM variance local to the source are 
depolarized below the detection limit at 144~MHz. With these sources missing from the LOFAR sample, our data are 
probing the low RM variance Universe, providing even more stringent constraints on the magnetization of the 
cosmic web away from galaxy cluster environments. 

To investigate the implication of our results for the strength of the co-moving cosmological magnetic field ($B_0$), 
we use a model of an inhomogeneous universe to calculate the RM difference between adjacent pairs of cosmic sight-lines. 
We use this model to generate \dRM~distributions for random and physical pairs for a wide range of input values of  
$B_0$ and the field correlation length. This allows us to place a limit of $B_0<4$~nG on Mpc scales. 

We also compare our results with a suite of cosmological MHD simulations, allowing us to investigate different 
magnetogenesis scenarios. In particular, we investigated the RM variance generated in three different scenarios: 
a strong initial primordial field of 0.5~nG, a weak primordial field of $10^{-9}$~nG but with dynamo amplification, 
and an astrophysical scenario where magnetic field is injected solely by AGN and galactic outflows. 
To constrain the different simulation scenarios, we use the observed RM structure function of physical pairs on 
angular scales less than 1~arcmin, because these data should have the lowest RM variance contribution 
from the Milky Way (which is not included in the model). 
We find that both the astrophysical scenario and a primordial 
scenario (with a seed field of $B\lesssim0.5$~nG) are consistent with the current data. 
Interestingly, the dynamo amplification in cosmic filaments is disfavoured because the RM dispersion is 
much larger than the observed scatter in the LOFAR data. 

In the coming years, we will be able to significantly expand on the current sample, in addition to adding redshift 
information for the host galaxies of the radio sources. This will allow us to push into the sub-nG regime and 
further constrain both the origin and evolution of cosmic magnetic fields on large scales. 

\section*{Acknowledgments}
SPO and MB acknowledge financial support from the Deutsche Forschungsgemeinschaft (DFG) under grant BR2026/23. 
MB acknowledges support from the Deutsche Forschungsgemeinschaft under Germany's Excellence 
Strategy - EXC 2121 ``Quantum Universe'' - 390833306. 
FV and NTL acknowledge financial support from the Horizon 2020 programme under the ERC Starting Grant ``MAGCOW'', no. 714196.  
The ENZO (enzo-project.org) simulations used for this work were produced on the CSCS Supercomputer 
of ETHZ (Lugano) and on the Marconi Supercomputer at CINECA (Bologna), under project no.~INA17\_C4A28 with FV 
as PI.  FV gratefully acknowledge the usage of online storage tools kindly provided by the INAF Astronomical Archive 
(IA2) initiative (http://www.ia2.inaf.it). 
CS acknowledges support from the ERC-StG DRANOEL, n.~714245. 
LOFAR \citep{vanhaarlem2013} is the Low
Frequency Array designed and constructed by ASTRON. It has observing, data
processing, and data storage facilities in several countries, that are owned by
various parties (each with their own funding sources), and that are collectively
operated by the ILT foundation under a joint scientific policy. The ILT resources
have benefitted from the following recent major funding sources: CNRS-INSU,
Observatoire de Paris and Universit\'e d'Orl\'eans, France; BMBF, MIWF-NRW, MPG,
Germany; Science Foundation Ireland (SFI), Department of Business, Enterprise and
Innovation (DBEI), Ireland; NWO, The Netherlands; The Science and Technology
Facilities Council, UK; Ministry of Science and Higher Education, Poland.
Part of this work was carried out on the Dutch national e-infrastructure with the support of the SURF 
Cooperative through grant e-infra 160022 \& 160152. The LOFAR software and dedicated reduction 
packages on https://github.com/apmechev/GRID\_LRT were deployed on the e-infrastructure by the 
LOFAR e-infragroup, consisting of J. B. R. Oonk (ASTRON \& Leiden Observatory), A. P. Mechev (Leiden Observatory) 
and T. Shimwell (ASTRON) with support from N. Danezi (SURFsara) and C. Schrijvers (SURFsara).  
This research has made use of data analysed using the University of Hertfordshire high-performance computing facility
(\url{http://uhhpc.herts.ac.uk/}) and the LOFAR-UK computing facility located at the University of Hertfordshire 
and supported by STFC [ST/P000096/1]. SPO thanks Marcel van Daalen for feedback on some aspects of the 
Monte Carlo model for the RM pairs. 
This research made use of Astropy, a community-developed core Python package for astronomy \citep{astropy2013} 
hosted at http://www.astropy.org/, of Matplotlib \citep{hunter2007}, of APLpy \citep{aplpy2012}, an open-source astronomical 
plotting package for Python hosted at http://aplpy.github.com/, and of TOPCAT, an interactive graphical viewer and 
editor for tabular data \citep{taylor2005}. 
The authors thank the referee, Prof.~Lawrence Rudnick, for a helpful review. 

\bibliographystyle{mnras.bst}
\bibliography{igmf} 

\begin{table}
\centering
\caption{Table of the coordinates, angular separation and RM values of all sources in the sample. 
The ID column indicates classification as a random or physical pair with the `r' or `p' suffix. 
The nominal RM error value does not include the error from the ionosphere RM correction, and 
thus is only valid in the case of taking the difference in RM between pairs in this catalog. }
\label{tab:data}
\begin{tabular}{lcccccc}
\hline
ID  & RA & Dec & $\Delta\theta$ & RM & RM error \\
   &  [J2000] & [J2000] & [arcmin] & [rad\,m$^{-2}$] & [rad\,m$^{-2}$] \\
\hline
1p & 00:18:09.27 & 31:01:19.19 & 2.48 & $-$76.848 & 0.045 \\
1p & 00:18:08.39 & 31:03:47.53 & 2.48 & $-$77.338 & 0.016 \\
2p & 00:29:00.04 & 29:42:15.88 & 1.22 & $-$62.950 & 0.010 \\
2p & 00:29:05.56 & 29:42:01.73 & 1.22 & $-$60.584 & 0.007 \\
3p & 00:44:34.09 & 12:11:26.59 & 0.80 & $-$15.581 & 0.013 \\
3p & 00:44:36.24 & 12:10:50.60 & 0.80 & $-$14.323 & 0.005 \\
4p & 00:45:59.20 & 22:26:54.03 & 7.50 & $-$45.683 & 0.016 \\
4p & 00:46:31.67 & 22:27:06.75 & 7.50 & $-$48.699 & 0.026 \\
5p & 00:46:54.30 & 12:57:06.82 & 3.56 & $-$12.789 & 0.006 \\
5p & 00:46:52.13 & 12:53:35.36 & 3.56 & $-$13.192 & 0.020 \\
6r & 00:47:06.84 & 12:44:52.99 & 12.61 & $-$11.804 & 0.030 \\
6r & 00:46:54.30 & 12:57:06.82 & 12.61 & $-$12.789 & 0.006 \\
7p & 00:51:02.21 & 13:13:37.38 & 5.52 & $-$14.196 & 0.054 \\
7p & 00:50:44.09 & 13:16:56.84 & 5.52 & $-$17.856 & 0.033 \\
8r & 00:53:23.20 & 33:27:25.21 & 8.87 & $-$57.024 & 0.016 \\
8r & 00:52:40.78 & 33:26:51.98 & 8.87 & $-$63.782 & 0.039 \\
9p & 01:01:23.40 & 29:28:52.50 & 1.85 & $-$67.635 & 0.043 \\
9p & 01:01:30.60 & 29:27:53.26 & 1.85 & $-$70.170 & 0.028 \\
\hline 
\end{tabular} \\
{Excerpt of the full table which is available online. }
\end{table}

\label{lastpage}
\end{document}